\documentclass[a4paper,fleqn,usenatbib]{mnras}

\usepackage{mathptmx}

\usepackage[T1]{fontenc}
\usepackage{ae,aecompl}

\usepackage{graphicx}   
\usepackage{amsmath}    
\usepackage{amssymb}    
\usepackage{longtable}
\usepackage[hyphenbreaks]{breakurl} 

\newcommand{\nova}{V906\,Car}

\DeclareSymbolFont{mysymbols}     {OMS}{cmsy}{m}{n}
\SetSymbolFont{mysymbols}  {bold}{OMS}{cmsy}{b}{n}
\DeclareSymbolFont{myoperators}   {OT1}{cmr} {m}{n}
\SetSymbolFont{myoperators}{bold}{OT1}{cmr} {bx}{n}
\DeclareMathSymbol{\forall}{\mathord}{mysymbols}{"38}
\DeclareMathSymbol{\exists}{\mathord}{mysymbols}{"39}
\DeclareMathSymbol{\pm}{\mathbin}{mysymbols}{"06}
\DeclareMathSymbol{+}{\mathbin}{myoperators}{"2B}
\DeclareMathSymbol{-}{\mathbin}{mysymbols}{"00}
\DeclareMathSymbol{=}{\mathrel}{myoperators}{"3D}
\DeclareMathSymbol{\times}{\mathbin}{mysymbols}{"02}

\newcommand*{\dittostraight}{---\textquotedbl---} 

\title[X-ray Spectroscopy of \nova{}]{X-ray Spectroscopy of the $\gamma$-ray Brightest Nova \nova{} (ASASSN-18fv)}

\author[K.~V.~Sokolovsky et al.]{
\parbox{\textwidth}{Kirill~V.~Sokolovsky$^{1,2,3}$\thanks{E-mail: kirx@kirx.net (KVS)},
Koji~Mukai$^{4}$,
Laura~Chomiuk$^{1}$,
Raimundo~Lopes de Oliveira$^{5,6}$,
Elias~Aydi$^{1}$,
Kwan-Lok~Li$^{7}$,
Elad~Steinberg$^{8}$,
Indrek~Vurm$^{9}$,
Brian~D.~Metzger$^{8}$,
Adam~Kawash$^{1}$,
Justin~D.~Linford$^{10}$,
Amy~J.~Mioduszewski$^{10}$,
Thomas~Nelson$^{11}$,
Jan-Uwe~Ness$^{12}$,
Kim~L.~Page$^{13}$,
Michael~P.~Rupen$^{14}$,\\
Jennifer~L.~Sokoloski$^{8}$,
Jay Strader$^{1}$}
\vspace{0.4cm}\\
\parbox{\textwidth}{
$^{1}$Center for Data Intensive and Time Domain Astronomy, Department of Physics and Astronomy, Michigan State University, 567 Wilson Rd, East Lansing, MI 48824, USA\\
$^{2}$Sternberg Astronomical Institute, Moscow State University, Universitetskii~pr.~13, 119992~Moscow, Russia\\
$^{3}$Astro Space Center of Lebedev Physical Institute, Profsoyuznaya~St.~84/32, 117997~Moscow, Russia\\
$^{4}$CRESST and X-ray Astrophysics Laboratory, NASA/GSFC, Greenbelt, MD 20771, USA\\
$^{5}$Departamento de F\'isica, Universidade Federal de Sergipe, Av. Marechal Rondon, S/N, 49000-000, S\~ao Crist\'ov\~ao, SE, Brazil\\
$^{6}$Observat\'orio Nacional, Rua Gal. Jos\'e Cristino 77, 20921-400, Rio~de~Janeiro, RJ, Brazil\\
$^{7}$Institute of Astronomy, National Tsing Hua University, Hsinchu 30013, Taiwan\\
$^{8}$Department of Physics and Columbia Astrophysics Laboratory, Columbia University, New York, NY 10027, USA\\
$^{9}$Tartu Observatory, University of Tartu, T\~oravere, 61602 Tartumaa, Estonia\\
$^{10}$National Radio Astronomy Observatory, Domenici Science Operations Center, 1003 Lopezville Road, Socorro, NM 87801, USA\\
$^{11}$Department of Physics and Astronomy, University of Pittsburgh, Pittsburgh, PA 15260, USA\\
$^{12}$XMM-Newton Observatory SOC, European Space Astronomy Centre, Camino Bajo del Castillo s/n, Urb. Villafranca del Castillo, E-28692 Villanueva de la Ca\~nada, Madrid, Spain\\
$^{13}$School of Physics and Astronomy, University of Leicester, University Road, Leicester, LE1 7RH, UK \\
$^{14}$National Research Council, Herzberg Astronomy and Astrophysics, 717 White Lake Rd, PO Box 248, Penticton, BC V2A 6J9, Canada}
}

\date{Accepted 2020 July 14. Received 2020 July 14; in original form 2020 April 13}

\pubyear{2020}

\begin{document}
\label{firstpage}
\pagerange{\pageref{firstpage}--\pageref{lastpage}}
\maketitle

\begin{abstract}
Shocks in $\gamma$-ray emitting classical novae are expected to produce bright thermal and non-thermal X-rays.
We test this prediction with simultaneous {\em NuSTAR} and {\em Fermi}/LAT observations of nova \nova{}, which exhibited 
the brightest GeV $\gamma$-ray emission to date. 
The nova is detected in hard X-rays while it is still $\gamma$-ray bright,
but contrary to simple theoretical expectations, the detected 3.5--78\,keV emission of \nova{} 
is much weaker than the simultaneously observed $>100$\,MeV emission. 
No non-thermal X-ray emission is detected, and our deep limits imply that the $\gamma$-rays are likely hadronic. 
After correcting for substantial absorption 
($N_\mathrm{H} \approx  2 \times 10^{23}$\,cm$^{-2}$), 
the thermal X-ray luminosity (from a 9\,keV optically-thin plasma) is just $\sim2$\% of the $\gamma$-ray luminosity.
We consider possible explanations for the low thermal X-ray luminosity, including 
the X-rays being suppressed by corrugated, radiative shock fronts or 
the X-rays from the $\gamma$-ray producing shock are hidden behind an even larger absorbing column ($N_\mathrm{H} >10^{25}$\,cm$^{-2}$).
Adding {\em XMM-Newton} and {\em Swift}/XRT observations to our analysis, we find that the evolution of the intrinsic X-ray absorption 
requires the nova shell to be expelled 24 days after the outburst onset. 
The X-ray spectra show that the ejecta are enhanced in nitrogen and oxygen, 
and the nova occurred on the surface of a CO-type white dwarf.
We see no indication of a distinct super-soft phase in the X-ray lightcurve, 
which, after considering the absorption effects, may point to a low
mass of the white dwarf hosting the nova. 
\end{abstract}




\section{Introduction}

\subsection{X-ray emission of classical novae}
\label{sec:xraynovaintro}

A nova explosion is powered by nuclear fusion that ignites at the bottom of 
a hydrogen-rich shell on the surface of an accreting white dwarf in a binary star system \citep{2008clno.book.....B,2016PASP..128e1001S}. 
Recent summaries of their observational appearance across the electromagnetic spectrum
were presented by \cite{2018arXiv180311529P} and \cite{2020arXiv200406540D}.
Specifically, the X-ray emission is produced during the following stages of a nova event \citep{2017PASP..129f2001M,2010AN....331..169H,2008clno.book.....K}:
\begin{enumerate}
 \item A soft X-ray flash should be produced by the optically-thick ejecta (i.e., ``fireball") during the first hours of explosion,
before the fireball expands and cools sufficiently to shift the peak of 
its emission from X-ray to UV and optical bands
\citep{2001MNRAS.320..103S,2002AIPC..637..345K,2007ApJ...663..505N}. 
So far, the attempts to observe the fireball X-ray emission have not resulted in an unambiguous detection \citep{2016PASJ...68S..11M,2016ApJ...830...40K}. 
\cite{2013ApJ...779..118M} interpret the X-ray transient MAXI\,J0158$-$744
as the nova fireball, while \cite{2012ApJ...761...99L} suggest the X-rays are
produced by interaction of the nova shell with the Be-type donor wind.
The ongoing all-sky X-ray surveys with 
MAXI/GSC \citep{2016PASJ...68S...1N} 
and 
SRG/eROSITA \citep{2012arXiv1209.3114M,2016SPIE.9905E..1KP} 
have a chance to detect a nova fireball.
 \item Hard X-ray emission ($\sim1-10$\,keV) is produced by optically-thin plasma
compressed and heated by internal shocks in the nova outflow and is often observed days to month 
after explosion \citep{1994MNRAS.271..155O,2014ASPC..490..327M}.
 \item Super-soft ($<0.5$\,keV, SSS) optically-thick thermal X-ray emission from 
the atmosphere of the nuclear-burning white dwarf is often observed when 
the nova ejecta become transparent to soft X-rays weeks-to-months after the explosion \citep{1994RvMA....7..129H,1997ARA&A..35...69K, 2011ApJS..197...31S}.
 \item Line-dominated emission from the shock-heated plasma may persist after the super-soft emission fades \citep{2014ATel.5920....1D,2009AJ....137.4627R}.
 \item When accretion restarts after the nova explosion, X-ray emission is produced in the region 
where accreting matter hits the white dwarf (the boundary layer between the disk and the surface 
in non-magnetic white dwarfs or the accreting column in magnetic ones). This is the accretion-powered X-ray emission found 
in cataclysmic variables (\citealt{2017PASP..129f2001M}, \citealt{2011PASJ...63S.729T}).
\end{enumerate}

The X-ray emission, including that powered by shocks, is widely assumed to be
thermal. However, detection of continuum GeV $\gamma$-ray emission from novae 
\citep{2010Sci...329..817A,2014Sci...345..554A,2016ApJ...826..142C,2018A&A...609A.120F} implies efficient particle 
acceleration by shocks \citep[e.g.][]{1994ApJS...90..515B,2014ApJ...783...91C,2015SSRv..188..187S}. 
The accelerated particles produce $\gamma$-rays through the 
hadronic (pion production and decay) and/or leptonic 
(direct acceleration of electrons 
and inverse Compton scattering of ambient photons or relativistic bremsstrahlung) mechanisms
\citep{2015MNRAS.450.2739M,2018A&A...612A..38M} -- the same mechanisms invoked
to explain high-energy emission from jetted active galactic nuclei known 
as ``blazars'' \citep{2010arXiv1006.5048B}.
If the hadronic scenario is responsible for the $\gamma$-ray production, 
novae should be sources of neutrinos that may be reachable for the next
generation detectors \citep{2016MNRAS.457.1786M}. In both hadronic and leptonic models,
the relativistic particles may
also contribute to non-thermal emission in X-rays \citep{2018ApJ...852...62V} and emit synchrotron
radiation in the radio band \citep{2016MNRAS.463..394V}.

It is also possible to produce non-thermal X-rays through Compton degradation of MeV line emission 
from decaying radioactive isotopes such as $^{22}$Na (\citealt{2010ApJ...723L..84S}, 
see also references in \citealt{2001MNRAS.326L..13O} and the discussion in \S\,\ref{sec:nustarspec}). 
The MeV line emission has long been predicted, but never observed \citep{2006NewAR..50..504H,2014ASPC..490..319H,2016stex.book.....J}.

The X-ray emission is absorbed by the expanding nova ejecta and, to an usually lesser extent, 
by the interstellar medium. The contributions of intrinsic and interstellar absorption may be disentangled as the intrinsic absorption decreases with time as the nova ejecta disperse \citep[e.g.][]{2001ApJ...551.1024M,2015MNRAS.454.3108P}.
The time it takes for the nova ejecta to thin out and reveal the underlying SSS may be used (together with the expansion velocity determined from optical spectroscopy) to estimate the nova ejecta mass \citep{2011ApJS..197...31S, 2014A&A...563A...2H}.

\subsection{X-rays from $\gamma$-ray detected novae}
\label{sec:xraynovagammaintro}
Surprisingly, no X-rays below 10\,keV (the energy band where most X-ray observatories, including {\em Swift/XRT}
and {\em XMM-Newton}, operate) have been observed from classical novae (i.e., novae with dwarf companions) 
while the novae were detected in $\gamma$-rays (\citealt{2014MNRAS.442..713M}, Gordon et~al.\ 2020, in prep.).
This might be explained if the soft X-rays are absorbed by the dense nova ejecta in the first weeks following explosion 
(see \S\,\ref{sec:ejectamass}). Interestingly, novae with red giant donors, like V407\,Cyg, are detected in X-rays 
simultaneously with the GeV emission, likely due to the shock being external---between the ejecta 
and the giant companion's wind---rather than internal to the nova ejecta 
\citep{2012ApJ...748...43N, 2012MNRAS.419.2329O}.

Thanks to its high sensitivity above 10\,keV, {\em NuSTAR} (\S\,\ref{sec:missionsintro}) can penetrate dense nova ejecta 
and constrain the X-ray luminosity simultaneously with the GeV detection by {\em Fermi}/LAT. 
V339\,Del and V5668\,Sgr were the first classical novae observed with {\em NuSTAR} 
while they were still bright in GeV $\gamma$-rays (\S\,\ref{sec:lxlg}); 
contrary to expectation, both resulted in non-detections (Mukai~et~al.\ 2020, in prep.). 
The first detection of X-rays simultaneous with $\gamma$-rays for a classical nova finally came with V5855\,Sgr, 
but deepened the mystery of the missing X-rays \citep{2019ApJ...872...86N}. Observed 12 days after eruption, 
the X-rays were consistent with highly-absorbed thermal plasma 
\citep[see e.g.][]{2013LNP...873.....G}, 
and the ratio of unabsorbed X-ray to $\gamma$-ray luminosity 
was $L_{\rm 20\,keV}/L_{\rm 100\,MeV} \approx 0.01$ 
(monochromatic flux ratio in $\nu F_{\nu}$ units; \S~\ref{sec:lxlg}). 
This ratio was surprisingly low, because we expect only a small fraction 
($\lesssim$10\%) of the shock energy to be transferred to the $\gamma$-ray emitting non-thermal particles
\citep[see \S\,\ref{sec:thermalx} and][]{2015MNRAS.450.2739M}. 
Meanwhile, the shocks in novae are predicted to be dense and radiative, implying that the bulk of the shock energy should 
be efficiently radiated away, and the shock speeds of $\gtrsim$1000\,km\,s$^{-1}$ imply that the bulk of this thermal emission 
should emerge in the X-ray band \citep{2014MNRAS.442..713M,2015MNRAS.450.2739M}. A possible explanation for the low value 
of $L_{\rm 20\,keV}/L_{\rm 100\,MeV}$ in V5855\,Sgr is suppression of X-rays at nova shock fronts \citep[][]{2019ApJ...872...86N}. 
If the shocks are dense and radiative, the shock front becomes subject to instabilities and develops a corrugated structure 
that can lead to post-shock temperatures a factor of $4-36$ lower than expected \citep{2018MNRAS.479..687S}. 
In that case, the shock luminosity is expected to emerge at longer wavelengths 
(i.e., optical/infrared) \citep{2014MNRAS.442..713M,2019arXiv190801700S}. Notably, correlated variations between the optical 
and $\gamma$-ray lightcurves of novae have now been observed in two
(possibly three, \citealt{2017MNRAS.469.4341M}) systems---including the subject of this paper, 
\nova{}---supporting this model (\citealt{2017NatAs...1..697L, 2020NatAs.tmp...79A}). 

Was the low value of $L_{\rm 20\,keV}/L_{\rm 100\,MeV}$ in V5855~Sgr unusual amongst novae? 
We know that the $\gamma$-ray properties of novae are diverse, with $>$100 MeV luminosities spanning at least 
a factor of $\sim$30 \citep{2018A&A...609A.120F}. And yet, we do not understand the cause of this diversity, 
or the full range of conditions in nova shocks. These open questions led us to observe \nova{} with {\em NuSTAR} 
while it was detected with {\em Fermi}/LAT---the results of which we present here.

\subsection{Orbital observatories}
\label{sec:missionsintro}

Our current understanding of nova X-ray emission comes primarily from 
{\em XMM-Newton} \citep[e.g.][]{2005ASPC..330..447H}
and {\em Swift} \citep{2007ApJ...663..505N,2011ApJS..197...31S,2012BASI...40..353N}
observations. {\em NuSTAR} has high sensitivity to very hard
X-rays and is just starting to reveal the behaviour of novae above 10\,keV.
{\em Fermi}/LAT detection of GeV emission from V407\,Cyg \citep{2010Sci...329..817A}
sparked a renewed interest in novae.
Here we briefly summarize the technical capabilities of these space missions.

The {\em Nuclear Spectroscopic Telescope Array (NuSTAR)}
\citep{2013ApJ...770..103H} was launched into a low-Earth orbit on 2012 June~13, 
equipped with two identical focusing X-ray telescopes sensitive to photons 
with energies 3--79\,keV \citep{2015ApJS..220....8M}. It provides two
orders of magnitude higher sensitivity (and an order of magnitude higher
angular resolution) compared to the coded aperture mask instruments
{\em Swift}/BAT and {\em INTEGRAL} sensitive to this energy range.
Its exceptional sensitivity makes new classes of objects, 
including classical novae, accessible for study in the hard X-ray regime.
The 10\,m-long extendible mast separating the X-ray optics and detector units
limits the speed at which the observatory can repoint, so  {\em NuSTAR} performs long observations of a single field interrupted by
Earth occultations, before repointing to another field (much like the 
{\em Hubble Space Telescope}).

The {\em Neil Gehrels Swift Observatory (Swift)} has been operating in 
low-Earth orbit since 2004 November~20 \citep{2004ApJ...611.1005G}. 
While originally designed for observations of $\gamma$-ray bursts and their afterglows, it became an
essential tool for multiwavelength studies across various branches of astronomy.
Its unique ability to quickly repoint makes it practical to perform
monitoring observations of multiple sources and 
allows {\em Swift} to use efficiently its time 
for observations (except for the South Atlantic Anomaly passages),
switching to a new target when the previous one goes into Earth occultation.
{\em Swift} is equipped with the coded aperture mask, wide field-of-view Burst
Alert Telescope \citep[BAT;][]{2005SSRv..120..143B} collecting very hard 15--150\,keV X-rays, 
the focusing X-ray Telescope \citep[XRT;][]{2005SSRv..120..165B} being sensitive to 0.3-10\,keV X-rays,  
and the Ultra-Violet/Optical Telescope \citep[UVOT;][]{2005SSRv..120...95R}.

{\em XMM-Newton} was launched on 1999 December~10 into a highly elliptical orbit allowing for 
long uninterrupted observations \citep{2001A&A...365L...1J}. 
The observatory can perform high-resolution X-ray grating spectroscopy
in the range 0.33--2.1\,keV (6--38\,\AA) using its Reflection Grating Spectrometer (RGS) instruments \citep{2001A&A...365L...7D}. 
It is also equipped for traditional medium-resolution
spectroscopy with a pair of EPIC-MOS\footnote{European Photon Imaging Camera - Metal Oxide Semiconductor \citep{2001A&A...365L..27T}} 
and the EPIC-pn\footnote{European Photon Imaging Camera with the pn-type detector \citep{2001A&A...365L..18S}.}  
imaging cameras covering a wider energy range of 0.2--10\,keV.
The EPIC-MOS and EPIC-pn cameras differ in the detector array geometry, electronics 
(resulting in different readout times) and quantum efficiency (front- and back-illuminated design, respectively).
The observatory also operates the Optical Monitor telescope \citep{2001A&A...365L..36M} 
that is similar to the {\em Swift}/UVOT. All {\em XMM-Newton} instruments are normally
operating simultaneously (the photons not intercepted by the RGS gratings
are collected by the EPIC-MOS cameras, while the EPIC-pn camera is fed by
its own X-ray mirror assembly).

{\em Fermi} Gamma-ray Space Telescope was launched into low-earth orbit on 2008 June~11.
Its main instrument, the Large Area Telescope 
(LAT; \citealt{2009ApJ...697.1071A,2009APh....32..193A,2012ApJS..203....4A})
is a pair-conversion detector sensitive to $\gamma$-rays in the energy range
20\,MeV--300\,GeV. Its collecting area and 2.4\,sr field of view are far
superior to the contemporary GeV telescopes AGILE \citep{2009A&A...502..995T,2008NIMPA.588...52T} 
and DAMPE \citep{2017APh....95....6C} and their predecessor -- EGRET, 
the spark chamber detector on board the {\em Compton} Gamma Ray Observatory
\citep{1993ApJS...86..629T}. {\em Fermi}/LAT normally performs an all-sky
survey every day, but see \S\,\ref{sec:latobs}.

\begin{table*}
        \centering
        \caption{{\em NuSTAR} observing log}
        \label{tab:nustarlog}
        \begin{tabular}{ccc cc ccc} 
                \hline
ObsID       & Epoch  & Start            & Stop             & Exposure  & Exposure  & Net count rate & Net count rate  \\
            & (days) &  UT              &  UT              & FPMA (ks) & FPMB (ks) & FPMA (cts/s)   & FPMB (cts/s)    \\
                \hline
80301306002 & 36 & 2018-04-20 14:46 & 2018-04-22 02:01 & 48.8      & 48.5      & $0.0158 \pm 0.0007$ & $0.0163 \pm 0.0007$ \\
90401322002 & 57 & 2018-05-11 16:26 & 2018-05-12 18:01 & 47.5      & 47.4      & $0.0434 \pm 0.0010$ & $0.0418 \pm 0.0010$ \\
                \hline
        \end{tabular}
\begin{flushleft}
{\bf Column designation:}
Col.~1~-- observation identification number;
Col.~2~-- time since outburst;
Col.~3~and~4~-- start and stop time of the observation (interrupted by Earth
occultations and South Atlantic Anomaly passes);
Col.~5~and~6~-- total on-source exposure time for FPMA and FPMB,
respectively;
Col.~7~and~8~-- source count rate (background-subtracted) for FPMA and FPMB,
respectively.
\end{flushleft}
\end{table*}

\subsection{\nova{} (2018)}
\label{sec:discoveryinfo}

\nova{} (Nova~Carinae~2018, ASASSN-18fv) was discovered on
2018 March~20.32\,UT \citep{2018ATel11454....1S} by the ASAS-SN survey
\citep{2014ApJ...788...48S,2017PASP..129j4502K}
as a new saturated object (<10\,mag) near the Carina Nebula. 
No previous outbursts were found by ASAS-SN or 
among the numerous amateur images of the Carina region \citep{2018MNSSA..77...25T}.
The initial spectroscopic observations by \cite{2018ATel11456....1S} on March~21 
and \cite{2018ATel11468....1I} on March~22 were unable to distinguish
between the possibilities of the object being a classical nova, a luminous
red nova \citep[e.g.][]{2019arXiv190600812P}, or a young stellar object outburst
\citep[e.g.][]{1996ARA&A..34..207H}. The main source of confusion were 
the low velocities derived from the emission lines at early times.
\cite{2018ATel11460....1L} obtained another spectrum on March~21 and interpreted it as that of a classical nova in the iron curtain phase. 
The infrared spectrum obtained by \cite{2018ATel11506....1R} on April~1 was consistent with
a Fe\,II-type nova, according to the classification scheme of \cite{1992AJ....104..725W}.

By a lucky coincidence, \nova{} was within the field of view of the
BRITE cubesat constellation \citep{2014PASP..126..573W,2016PASP..128l5001P,2017A&A...605A..26P}, 
as it was performing photometry of a nearby red giant HD\,92063 
(see also \S\,\ref{sec:xmm} and \ref{sec:swiftmonitoring};
\citealt{2018ATel11508....1K}).
We adopt $t_0=$ 2018 March~16.13~UTC {(\rm HJD}2458193.63) as the nova explosion
time derived from the BRITE lightcurve by 
\cite[][see their Supplementary Figure~2]{2020NatAs.tmp...79A}. The adopted $t_0$ is
consistent with the reported non-detection by Evryscope two hours earlier \citep{2018ATel11467....1C}. 
The optical lightcurve of \nova{}, peaking at 5.9\,mag, showed an unusual series of 
fast flares superimposed on the slowly evolving nova lightcurve. 

As of 2020 June, \nova{} is the brightest $\gamma$-ray nova observed by {\em Fermi}/LAT
to date \citep{2018ATel11546....1J}, reaching peak 0.1--300\,GeV flux of 
$(1.91 \pm 0.20) \times 10^{-6}$~photons~cm$^{-2}$~s$^{-1}$ in a 12\,h integration centred on 
2018 April~14.25~UT ($t_0+29$\,days; \S\,\ref{sec:latobs}; \citealt{2020NatAs.tmp...79A}).
Remarkably, a series of distinct flares was resolved in the {\em Fermi}/LAT
lightcurve that coincided with the optical flares observed by BRITE. 
This led \cite{2020NatAs.tmp...79A} to conclude that these flares are manifestations of shocks. 
\nova{} was also the first nova detected by 
the {\em AGILE} mission \citep{2018ATel11553....1P} observing at the $>100$\,MeV band similar to {\em Fermi}/LAT.

\nova{} was observed by {\em INTEGRAL} starting on 2018 April~23
($t_0+38$\,days), with the aim of searching for the MeV $\gamma$-ray nucleosynthesis 
lines predicted in novae\footnote{\url{https://www.cosmos.esa.int/web/integral/news-2018}}, 
one of the long-standing goals of the {\em INTEGRAL} mission
\citep{2002AIPC..637..435H,2018A&A...615A.107S}. 
No MeV line detections were reported. High-cadence optical photometry
was obtained with {\em INTEGRAL}/OMC, revealing variations of up to 0.3\,mag
on time-scales of several hours to one day \citep{2018ATel11677....1D}.

\cite{2020MNRAS.494..743M} report dense monitoring
of \ion{Fe}{II} and \ion{[O}{I]} features in the optical spectrum of
\nova{}. The authors argue that these spectral features might be originating
in a rotating circumbinary disc. \cite{2020MNRAS.495.2075P} also report
spectroscopy of \nova{} concluding that the nova ejecta are clumpy and have an
overall asymmetric bipolar geometry.

\nova{} was also observed at radio wavelengths with the Australia Telescope Compact Array
(ATCA) resulting in an initial non-detection on 2018 April~03
\citep[$t_0+18$\,days;][]{2018ATel11504....1R}. The mJy-level radio emission was
detected first on 2018 May~13 ($t_0+58$\,days) and 
reached its peak in late 2019 \citep{2020NatAs.tmp...79A}.

Hard X-ray emission from \nova{} was detected by {\em NuSTAR}, 
and preliminary results were reported in \citet{2018ATel11608....1N} and \citet{2020NatAs.tmp...79A}. 
Here we present a more in-depth look at the X-ray emission from this nova,
analyzing the {\em NuSTAR} data together with {\em XMM-Newton} and 
{\em Swift} observations.

\subsection{Galactic extinction towards \nova{}}
\label{sec:extinction}

Optical spectroscopy allowed \citet{2020NatAs.tmp...79A} to
estimate the interstellar reddening towards \nova{} using the diffuse
interstellar bands \citep{2011ApJ...727...33F} and the \ion{Na}{i}~D
 absorption features \citep{2012MNRAS.426.1465P}. Combining these two methods, 
the authors found $E(B-V)=0.36 \pm 0.05$.
Assuming the standard value of $R_V=3.1=A_V/E(B-V)$ \citep{1975A&A....43..133S}, 
this corresponds to a $V$-band extinction of $A_V=1.12$\,mag. 

An alternative method of estimating extinction to a nova is based on 
the typical intrinsic colour of $(B-V)_0=-0.02$ when the nova is two magnitudes 
below its peak \citep[the dispersion of $(B-V)_0$ is $0.12$\,mag,][]{1987A&AS...70..125V}.
The nova light at this stage may be dominated by optically-thick free-free
(blackbody) emission \citep{2014ApJ...785...97H}. 
According to \cite{2020NatAs.tmp...79A}, for \nova{} the observed $(B-V)=0.23$ around
$t_0+55$\,days, corresponding to a colour excess of $E(B-V)=0.25$ -- consistent
with the spectroscopically-derived value within one sigma uncertainty of the intrinsic color.
We adopt the spectroscopically-derived $E(B-V)$ as it has lower uncertainty.
It is also not clear whether the method based on intrinsic colour is applicable to \nova{}, 
considering the major contribution of shock-powered optical emission \citep{2020NatAs.tmp...79A}.

We note that \cite{2020MNRAS.495.2075P}, relying on a different \ion{Na}{i}~D 
equivalent width--reddening calibration and the expected nova color around
maximum light, arrive at a much higher $E(B-V)$ value. This value, however,
would imply the Galactic X-ray absorbing column is much higher than that
we derive from {\em XMM-Newton} observations, as described in \S~\ref{sec:xmm}.

To estimate the expected Galactic X-ray absorbing column to \nova{}, 
we utilize the relation proposed by \cite{2009MNRAS.400.2050G} between 
the optical extinction and the hydrogen column density:
\begin{equation}
N_\mathrm{H} = 2.21 \times 10^{21}\,{\rm cm}^{-2} \times A_V = 2.47 \times 10^{21}\,{\rm cm}^{-2}
\end{equation}
This value should be a lower limit on the total 
X-ray absorbing column, as the nova ejecta produce large intrinsic
absorption (\S\,\ref{sec:ejectamass}). The derived $N_\mathrm{H}$ is
consistent with the value derived from our late-time {\em XMM-Newton}
spectroscopy (Table~\ref{tab:xmm}).
The total \ion{H}{I} column density in the direction of \nova{} derived 
from Galactic surveys of the 21\,cm emission line is 
$N_\mathrm{HI} = 1.29 \times 10^{22}$\,cm$^{-2}$ \citep{2005A&A...440..775K,2005A&A...440..767B}, 
so the nova is in front of 80\% of the Galactic absorbing column. 
Comparison of the $A_V$ estimated for \nova{} to 
the total optical V-band extinction in its direction \citep[3.6\,mag;][]{2011ApJ...737..103S}
also suggests that the nova is nearby. \cite{2020NatAs.tmp...79A} adopted a
distance to \nova{} of 4\,kpc based on the uncertain {\em Gaia} parallax
measurement \citep{2018AJ....156...58B} and 
the Galactic 3D extinction map of \cite{2019MNRAS.483.4277C}.

\subsection{Scope of this work}
\label{sec:thispaper}

We present a joint analysis of {\em NuSTAR}, {\em XMM-Newton}, {\em Swift}
and {\em Fermi} observations of \nova{}. 
We fit the model describing the observed X-ray spectra and constrain the elemental abundances in the nova ejecta. 
We compare the simultaneous hard X-ray ({\em NuSTAR}) and $\gamma$-ray ({\em Fermi}/LAT) 
observations to identify the physical origin of the high-energy emission in this
nova, and discuss the possible location of the X-ray emitting shock.
In Section~\ref{sec:obs}, we describe the observations of \nova{} performed with the
instruments introduced in \S\,\ref{sec:missionsintro}. In Section~\ref{sec:discussion},
we estimate physical parameters of the nova, and summarize our findings in
Section~\ref{sec:conclusions}.

Throughout this paper we adopt a significance level $\alpha_{\rm lim} = 0.05$: 
we reject spectral models that have a probability $p>\alpha_{\rm lim}$ of obtaining the observed or 
a more extreme value of the test statistic 
by chance\footnote{see e.g. Chapter~5 of \cite{2003psa..book.....W} and \url{https://en.wikipedia.org/wiki/P-value}}. 
We use $\chi^2$ as the test statistic as we deal with well-sampled spectra (\S\,\ref{sec:nustarspec}). 
We express the abundances of the chemical elements by the number of atoms relative 
to the number of hydrogen atoms following the \texttt{XSPEC} 
convention (\S\,~\ref{sec:ejectaabund}). 
The quoted uncertainties of the model parameters are at $1 \sigma$ level.
For power law emission, we define the spectral index $\alpha$ as 
$F_\nu \propto \nu^\alpha$ 
where $F_\nu$ is the spectral flux density and $\nu$ is the frequency; 
meanwhile the corresponding index in the distribution of the number of incoming photons as 
a function of energy is $dN(E)/dE \propto E^{-\Gamma}$, 
where $\Gamma$ is called the photon index and $\Gamma = 1 - \alpha$.
The same power law expressed in spectral energy distribution units 
(SED; \citealt{1997NCimB.112...11G}) is 
$\nu F_\nu \propto \nu^{\alpha + 1} \propto \nu^{-\Gamma + 2}$.
When referring to ``GeV $\gamma$-rays'' we imply emission in 
the {\em Fermi}/LAT band (0.1--300\,GeV).

\begin{figure*}
\begin{center}
\includegraphics[height=0.33\textwidth,clip=true,trim=0cm 0.5cm 0.5cm 0.5cm,angle=0]{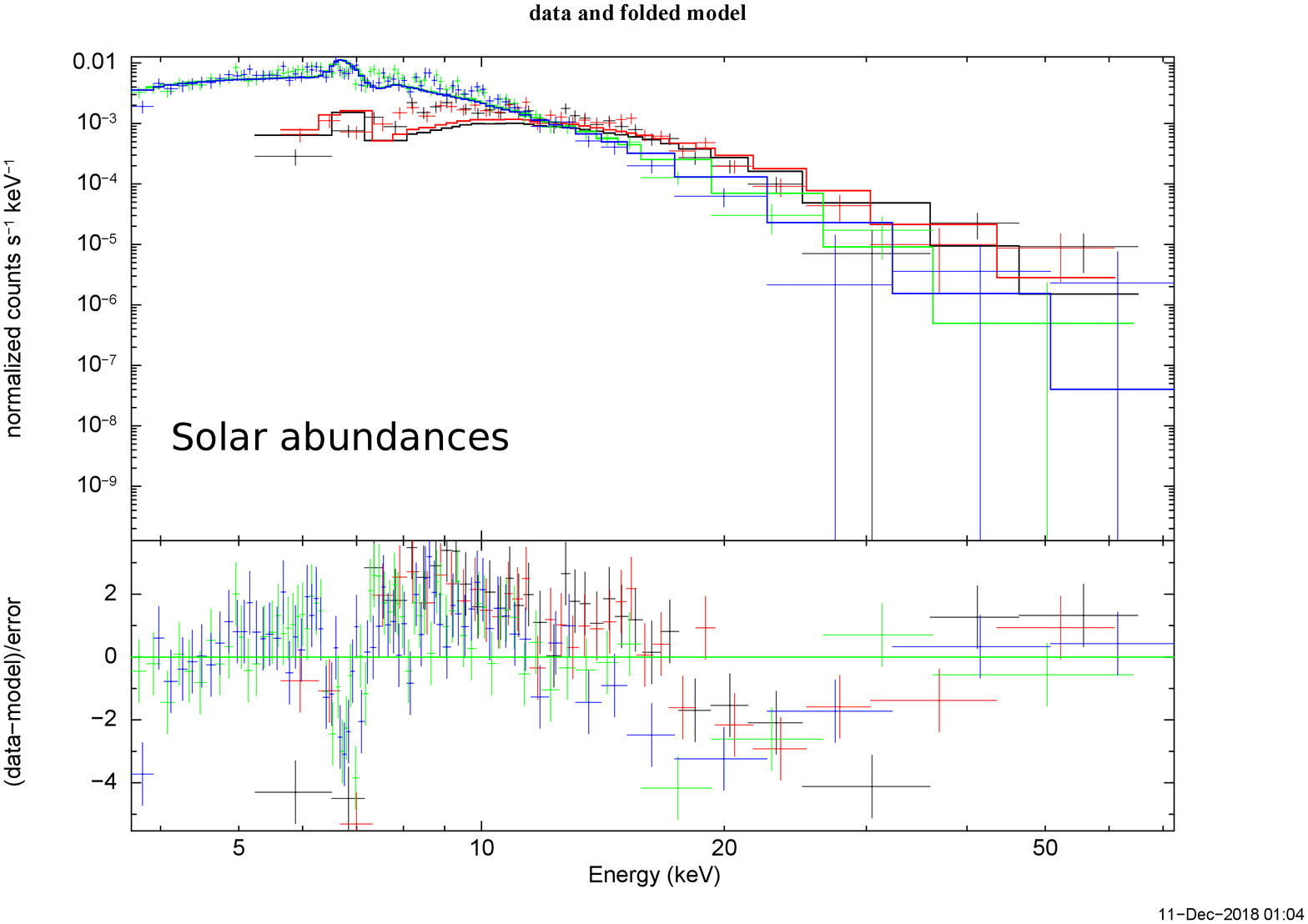}
\includegraphics[height=0.33\textwidth,clip=true,trim=0cm 0.5cm 0.5cm 0.5cm,angle=0]{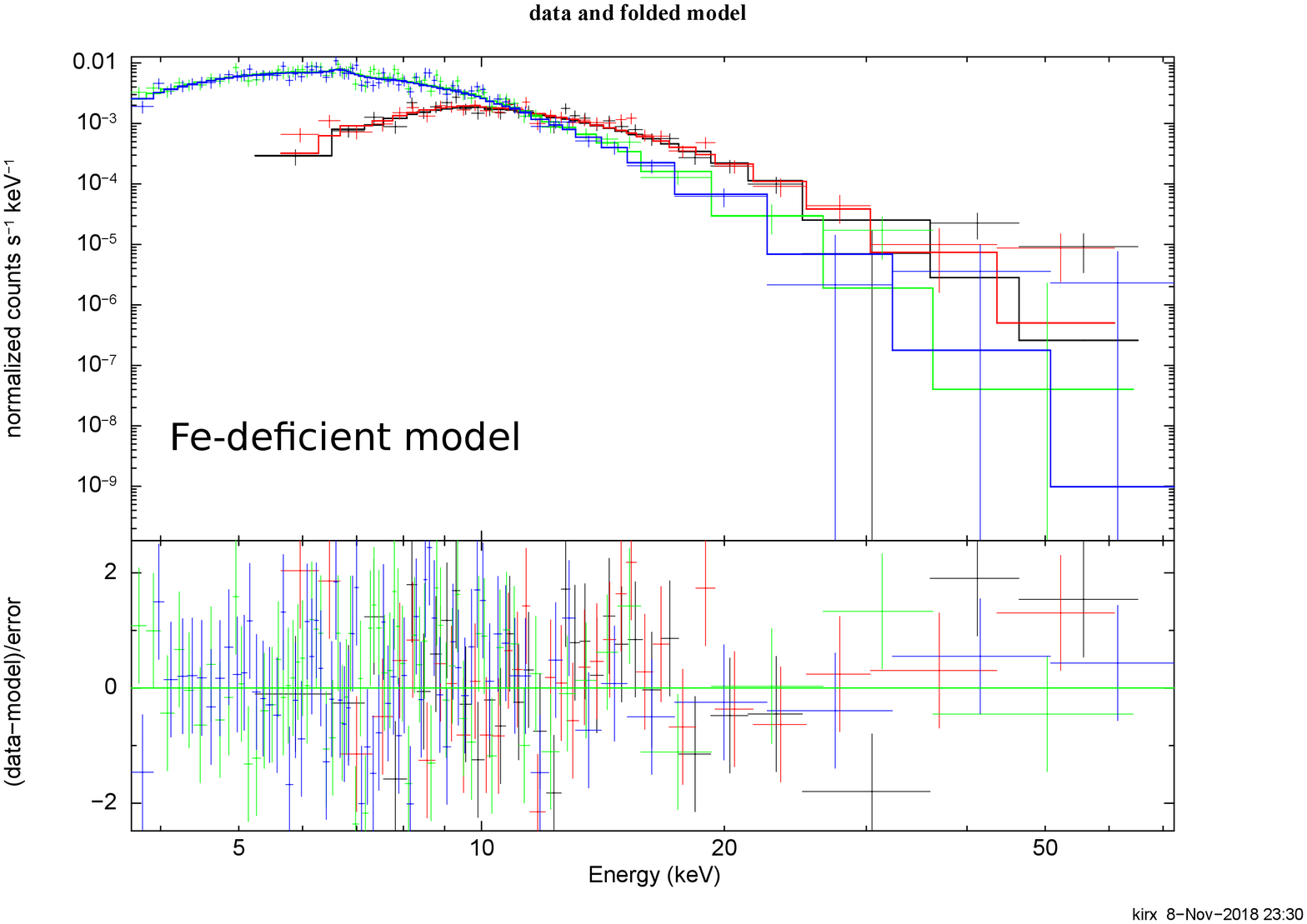}
\includegraphics[height=0.33\textwidth,clip=true,trim=0cm 0.5cm 0.5cm 0.5cm,angle=0]{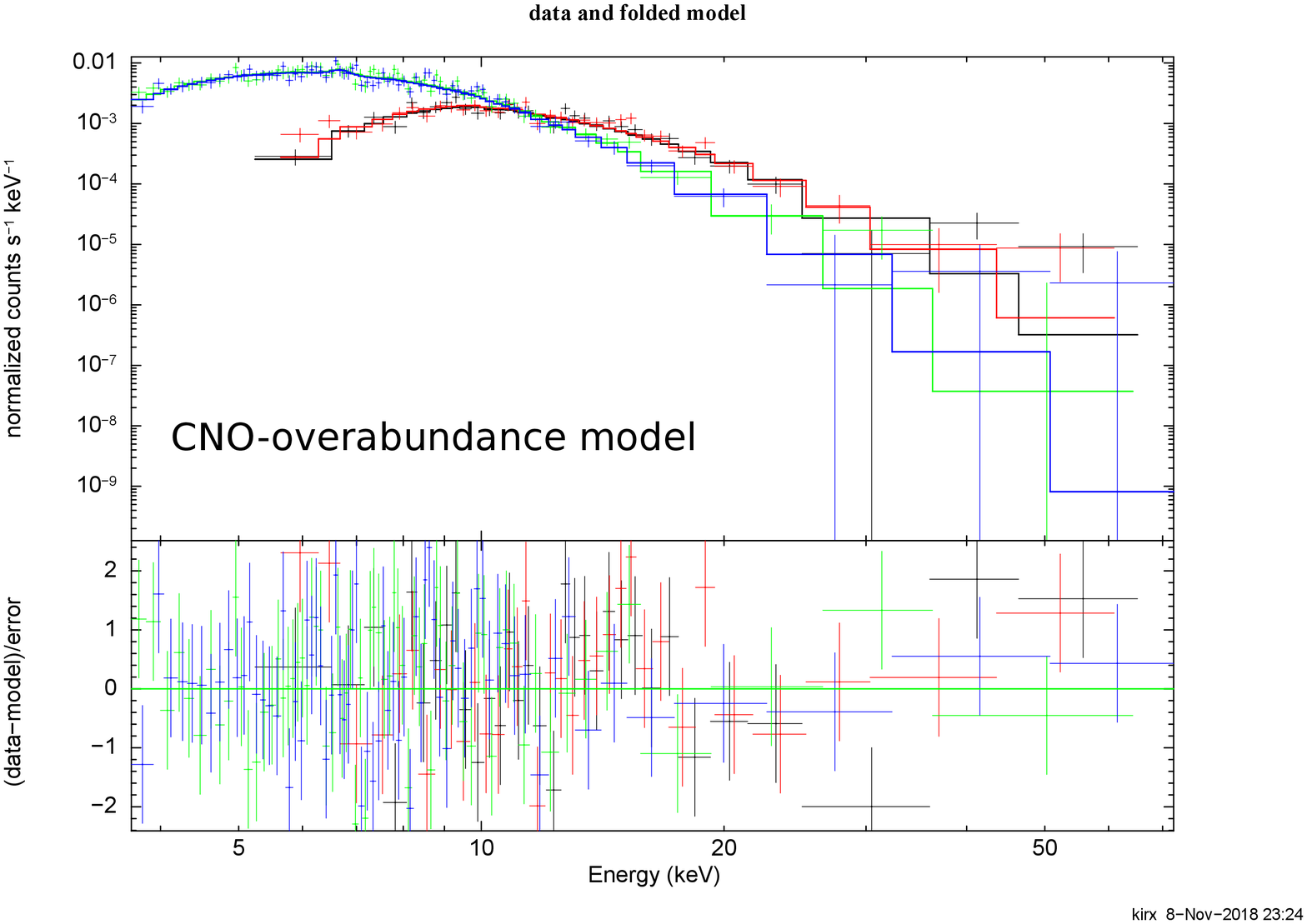}
\includegraphics[height=0.33\textwidth,clip=true,trim=0cm 0.5cm 0.5cm 0.5cm,angle=0]{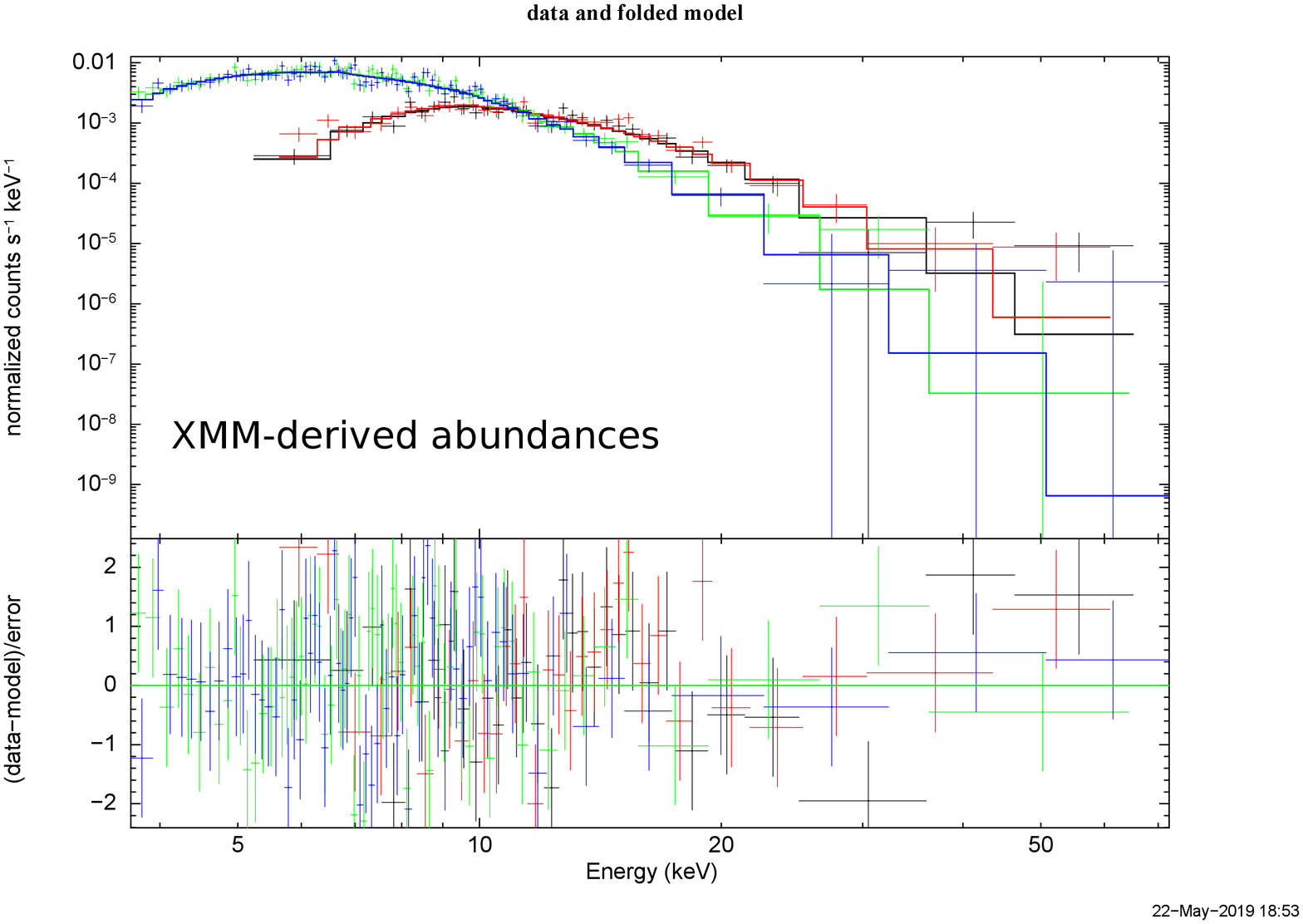}
\end{center}
\caption{Observed {\em NuSTAR} spectra compared with four different models for absorbed thermal plasma
emission model: abundances fixed to solar (top left panel);
FeCoNi abundances tied together and left free to vary (top right panel);
CNO abundances tied together and left free to vary, while FeCoNi are fixed to solar values (bottom
left panel); 
and the CNOFeCoNi abundances fixed to the values derived from the {\em XMM-Newton} observations (bottom right panel; see Table~\ref{tab:xmm}). 
Only the latter three models produce statistically acceptable fits (Table~\ref{tab:nustarmodels}).
Black and red curves represent spectra obtained with two {\em NuSTAR} telescopes FPMA and FPMB during the first epoch ($t_0+36$\,d), while green and
blue curves are the FPMA and FPMB spectra obtained during the second epoch ($t_0+57$\,d).}
\label{fig:nuspec}
\end{figure*}

\section{Observations and analysis}
\label{sec:obs}

\subsection{{\em NuSTAR} spectroscopy}
\label{sec:nustarspec}

{\em NuSTAR} observed \nova{} during two epochs: $t_0+36$ and $t_0+57$\,d. 
The nova is clearly detected with 1563 and 4046 counts in the source region 
(two focal plane modules combined) in the first and second epoch, respectively.
The observing log is presented in Table~\ref{tab:nustarlog}.
We use \texttt{nupipeline} and \texttt{nuproducts} commands from
\texttt{HEASoft\,6.26.1} to extract source and background spectra from 
the focal plane modules A (FPMA) and B (FPMB). 
A circular extraction region with radius of 30\,$\arcsec$
was centred on the X-ray image of the nova using \texttt{ds9} \citep{2003ASPC..295..489J}.
The centring was done for FPMA and FPMB event files separately. 
The background was extracted from five circular regions of the same size 
that were manually placed near the nova on the same 
CZT \citep{2011hxra.book.....A} chip. 
We checked that the specific choice of the background region does not affect the results.
We use \texttt{grppha} to 
mark channels 0--46 and 1910--4095 as ``bad'', restricting the energy range to 
3.5--78.0\,keV and grouping the source spectra to contain at least 25 counts per bin.
The spectra together with the redistribution matrix (RMF; describes the
probability of a count being registered at a certain energy channel as 
a function of the photon energy) and auxiliary response
(ARF; describes the detector effective area as a function of energy) calibration 
files provided by the pipeline are loaded into \texttt{XSPEC\,12.10.0c} \citep{1996ASPC..101...17A} for further analysis.

\begin{table*}
        \centering
        \caption{Parameters of {\em NuSTAR} spectral fits}
        \label{tab:nustarmodels}
        \begin{tabular}{c@{~~~}c@{~~~}c@{~~~}c@{~~~}c@{~~~}c@{~~~}c@{~~~}c} 
                \hline
$t-t_0$ & \texttt{vphabs} $N_\mathrm{H}$ & k$T$  & FeCoNi                  & CNO                  &  $C_{\rm FPMB}$    & 3.5--78.0\,keV Flux             & 3.5--78.0\,keV Flux$_{0}$\\ 
(days)  & ($10^{22}$\,cm$^{-2}$)         & (keV) & abundances              & abundances           &                    & $\log_{10}$(erg\,cm$^{-2}$\,s$^{-1}$)  & $\log_{10}$(erg\,cm$^{-2}$\,s$^{-1}$) \\
                \hline
\multicolumn{8}{c}{Solar abundances model: 
$\chi_{\rm red}^2=3.1047$, ${\rm d.o.f.}=200$, $p=0.00$} \\
36    &  $165 \pm 14$         & $13.7 \pm 1.7$  & 1.0                      & 1.0                   & $1.23 \pm 0.08$  &  $-11.570 \pm 0.012$             & $-11.068 \pm 0.012$    \\
57    &  $16.4 \pm 1.6$       & $7.5 \pm 0.3$   & \dittostraight            & \dittostraight       & $1.01 \pm 0.04$  &  $-11.454 \pm 0.007$             & $-11.179 \pm 0.007$    \\
\\
\multicolumn{8}{c}{Fe-deficient model: 
$\chi_{\rm red}^2=1.0281$, ${\rm d.o.f.}=199$, $p=0.38$} \\
36    &  $293 \pm 20$         & $8.0 \pm 0.9$  & $0.09 \pm 0.03$ & 1.0                   & $1.11 \pm 0.06$  &  $-11.570 \pm 0.012$             & $-11.068 \pm 0.012$    \\
57    &  $44.8 \pm 2.7$       & $4.4 \pm 0.2$  & \dittostraight            & \dittostraight        & $1.01 \pm 0.03$  &  $-11.454 \pm 0.007$             & $-11.179 \pm 0.007$    \\
\\
\multicolumn{8}{c}{CNO-overabundance model: 
$\chi_{\rm red}^2=1.0457$, ${\rm d.o.f.}=199$, $p=0.31$} \\
36    & $4.3 \pm 2.3$         & $8.6 \pm 0.9$ & 1.0                       & $210 \pm 110$     & $1.11 \pm 0.06$  &  $-11.564 \pm 0.012$             & $-11.143 \pm 0.012$    \\
57    & $0.6 \pm 0.3$         & $4.4 \pm 0.2$ & \dittostraight            & \dittostraight        & $1.01 \pm 0.03$  &  $-11.454 \pm 0.007$             & $-11.221 \pm 0.007$    \\
\\
\multicolumn{8}{c}{{\em XMM}-derived abundances model: 
$\chi_{\rm red}^2=1.0552$, ${\rm d.o.f.}=200$, $p=0.28$} \\
36    & $19.3 \pm 1.3$        & $8.6 \pm 0.8$ & 0.10                      & ${\rm C}=0$, ${\rm O}=29$,                    & $1.10 \pm 0.06$  &  $-11.566 \pm 0.012$             & $-11.143 \pm 0.012$    \\
57    & $2.6 \pm 0.2$         & $4.3 \pm 0.2$ & \dittostraight            & ${\rm N}=345$                           & $1.01 \pm 0.03$  &  $-11.455 \pm 0.007$             & $-11.210 \pm 0.007$    \\
\\
\multicolumn{8}{c}{{\em XMM} abundances and fixed Galactic column model: 
$\chi_{\rm red}^2=1.0547$, ${\rm d.o.f.}=200$, $p=0.28$} \\
36    & $19.3 \pm 1.3$        & $8.6 \pm 0.8$ & 0.10                      & ${\rm C}=0$, ${\rm O}=29$,                    & $1.11 \pm 0.06$  &  $-11.566 \pm 0.012$             & $-11.141 \pm 0.012$    \\
57    & $2.64 \pm 0.16$       & $4.3 \pm 0.2$ & \dittostraight            & ${\rm N}=345$                           & $1.01 \pm 0.03$  &  $-11.455 \pm 0.007$             & $-11.209 \pm 0.007$    \\
\\
\multicolumn{8}{c}{Two-temperature plasma model: 
$\chi_{\rm red}^2=1.0292$, ${\rm d.o.f.}=196$, $p=0.37$} \\
36    & $27.6 \pm 4.3$        & $6.9 \pm 0.7$, $0.57 \pm 0.07$ & 0.10                      & ${\rm C}=0$, ${\rm O}=29$,              & $1.12 \pm 0.06$  &  $-11.582 \pm 0.012$             & $-9.624 \pm 0.012$    \\
57    & $3.5 \pm 1.5$         & $4.1 \pm 0.4$, $0.58 \pm 0.17$ & \dittostraight            & ${\rm N}=345$                           & $1.01 \pm 0.03$  &  $-11.456 \pm 0.007$             & $-11.074 \pm 0.007$    \\
\\
\multicolumn{8}{c}{Power-law model: 
$\chi_{\rm red}^2=1.2667$, ${\rm d.o.f.}=200$, $p=0.006$} \\
36    & $24.4 \pm 1.8$        & $\Gamma = 3.30 \pm 0.18$ & 0.10                      & ${\rm C}=0$, ${\rm O}=29$,              & $1.10 \pm 0.06$  &  $-11.500 \pm 0.012$              & $-10.833 \pm 0.012$    \\
57    & $4.5 \pm 0.2$         & $\Gamma = 3.92 \pm 0.10$ & \dittostraight            & ${\rm N}=345$                           & $1.01 \pm 0.03$  &  $-11.426 \pm 0.007$              & $-10.949 \pm 0.007$    \\
                \hline
        \end{tabular}
\begin{flushleft}
The preferred model is marked in boldface.
{\bf Column designation:}
Col.~1~-- observation time, in units of days since outburst;
Col.~2~-- equivalent hydrogen column density;
Col.~3~-- plasma temperature;
Col.~4~-- abundances of Fe, Co and Ni (tied together) relative to the solar values; 
Col.~5~-- abundances of C, N and O (tied together for the first two models) relative to the solar values;
Col.~6~-- normalization factor of FPMB relative to FPMA;
Col.~7~-- absorbed model flux in the energy range 3.5--78.0\,keV;
Col.~8~-- unabsorbed 3.5--78.0\,keV flux.
\end{flushleft}
\end{table*}

To fit the {\em NuSTAR} observations we first choose the absorbed 
optically-thin thermal equilibrium plasma model attenuated by photoelectric
absorption: \texttt{XSPEC} model \texttt{constant*vphabs*vapec}.
We simultaneously fit all four spectra (FPMA and FPMB spectra obtained at
two epochs) allowing for absorbing column (\texttt{vphabs}), 
temperature and normalization factor of \texttt{vapec}
variations between epochs. The normalization factor between the FPMA and FPMB spectra
(represented by the \texttt{constant} term) is also allowed to vary
between the two epochs. The best-fitting model parameters, together with their
estimated $1\sigma$ uncertainties, are listed in Table~\ref{tab:nustarmodels}.

The observed X-ray spectrum cannot be fit by an absorbed thermal plasma 
if we assume solar abundances 
(top left panel of Fig.~\ref{fig:nuspec}). 
The $\chi_{\rm red}^2$ value we find (3.1; Table~\ref{tab:nustarmodels}) corresponds to the
null hypothesis probability of $p << \alpha_{\rm lim}$. 
The data systematically depart from the model predictions around 6.7 and
20-30\,keV, which makes it even less likely to occur by chance
compared to the simple $\chi^2$ statistics that does not take into account
correlations in residuals \citep[c.f. the `alarm' statistic of ][]{2006MNRAS.367.1521T}.
Throughout this work we always assume the same abundances for the emitter and absorber
(with the exception of Galactic absorbing component that we consider separately below).

At least two variations of the absorbed thermal plasma model are 
compatible with the observations. The first is a model with Fe abundance (by number) of $0.09\pm0.03$
times the solar value (top right panel of Fig.~\ref{fig:nuspec}).
Fe is present in the nova ejecta (as we clearly see Fe lines in the optical spectrum; \citealt{2018ATel11460....1L}; \citealt{2020NatAs.tmp...79A}) 
but it may be under-abundant with respect to solar values.
The second model that provides a good fit to the \emph{NuSTAR} spectrum has solar Fe abundance and overabundance
of CNO elements by a factor of $210 \pm 110$ (bottom left panel of Fig.~\ref{fig:nuspec}).
Novae are known to show overabundance of CNO elements 
\citep[][and the discussion in \S\,\ref{sec:ejectaabund}]{1985ESOC...21..225W,1994ApJ...425..797L,1998PASP..110....3G,2001MNRAS.320..103S}.

The results in Table~\ref{tab:nustarmodels} do not depend strongly on the specific choice of solar abundances 
(we used the latest abundances from \citealt{2009ARA&A..47..481A} available in \texttt{XSPEC}, 
but also tested the values from \citealt{2000ApJ...542..914W} and \citealt{2003ApJ...591.1220L}). 
We found that our spectral fits minimally depend on the choice of absorption model, 
comparing \texttt{tbvarabs} \citep{2000ApJ...542..914W} to \texttt{vphabs} \citep{1992ApJ...400..699B}. 
We also fit an alternative thermal plasma emission model (\texttt{vmekal}; \citealt{1985A&AS...62..197M,1995ApJ...438L.115L}), 
and found the difference with the \texttt{vapec} model fit was within the statistical errors.
An acceptable fit ($\chi_{\rm red}^2=1.04$, ${\rm d.o.f.}=199$, $p=0.32$) can be achieved with 
the simple thermal bremsstrahlung model \texttt{bremss} \citep{1975ApJ...199..299K}, 
implying the absence of obvious emission features in the {\em NuSTAR} spectra. 
We prefer the \texttt{vapec} model \citep{2005AIPC..774..405B} over \texttt{bremss} as
this model is more physically motivated (we expect the line emission to be
present at some low level, see \S\,\ref{sec:xmm}).

We also fit the {\em NuSTAR} spectra with an absorbed thermal plasma
model (\texttt{vphabs*vapec}), fixing the abundances set to the values 
derived from our {\em XMM-Newton} observations described in \S\,\ref{sec:xmm} and Table~\ref{tab:xmm} 
(see the bottom right panel of Fig. \ref{fig:nuspec}). 
The C abundance that is not well constrained from the {\em XMM-Newton}
spectroscopy was set to the solar value. We checked that the fit remains essentially 
the same if we set the C abundance to 0. The Co and Ni abundances were set equal to Fe.
The resulting plasma temperature (k$T$) and unabsorbed flux are close to 
the ones suggested by the CNO-overabundance model, while
the absorbing column is best fit by a value intermediate between
the CNO-overabundance and Fe-deficient models (Table~\ref{tab:nustarmodels}).
To estimate the errors in k$T$ and $N_\mathrm{H}$ resulting from uncertainties in abundances, 
we vary the N and O abundances within the errors of the {\em XMM-Newton} spectrum fitting 
(Table~\ref{tab:xmm}), 
while for Fe we vary the abundances between 0.0 and 0.1 and C in the range 0.0--1.0. 
These input parameter variations result in best-fitting $N_\mathrm{H}$ values in the range $(15.4-23.5) \times 10^{22}$\,cm$^{-2}$ for the
first epoch and $(2.1-3.3) \times 10^{22}$\,cm$^{-2}$ for the second epoch (these are full ranges of the obtained best-fitting values, not confidence intervals).
The corresponding k$T$ range is 8.63--8.70\,keV and 4.32--4.35\,keV for 
the first and second epochs, respectively. 
Comparing the ranges of k$T$ and $N_\mathrm{H}$ values obtained with various abundances to the
best-fitting values and their uncertainties for the models listed in Table~\ref{tab:nustarmodels}, 
one can see that the temperatures are largely insensitive to the choice of abundances, 
while the $N_\mathrm{H}$ values strongly depend on that choice.

As the joint {\em NuSTAR} and {\em Swift}/XRT observations of the recurrent nova 
V745\,Sco by \cite{2015MNRAS.448L..35O} were fit with two-temperature plasma, 
we tried adding a second \texttt{vapec} component to our {\em NuSTAR} model 
(\texttt{constant*phabs*vphabs*vapec}; Table~\ref{tab:nustarmodels}; 
\texttt{phabs} component describes the fixed Galactic $N_\mathrm{H}$, as discussed below).
The fit suggests that a very bright component 
(0.3--78.0\,keV flux of $10^{-7}$\,erg~cm$^{-2}$~s$^{-1}$ and 
$7 \times 10^{-10}$\,erg~cm$^{-2}$~s$^{-1}$, for the first and second epochs, respectively) 
can be hidden behind the inferred absorbing column, 
if this second emission component is soft (k$T \simeq 0.6$\,keV).
The presence of this low-energy component is not required to obtain a statistically 
acceptable fit so its flux is highly uncertain and consistent with being
zero. The dramatic decrease in the best-fitting flux of this
hypothetical low-energy component between the two epochs likely reflects 
the improving constraints on the soft emission resulting from the decreasing
absorbing column, rather than any real change in the emission.
The addition of the soft \texttt{vapec} component does not change the
parameters of the hard \texttt{vapec} component; they remain essentially the
same as in Table~\ref{tab:nustarmodels} for the single-component emission. 

As many novae show super-soft emission at some point in their X-ray evolution (\S\,\ref{sec:xraynovaintro}, \S\,\ref{sec:missingsss}),
we also try to replace the low-energy \texttt{vapec} component with a black
body. This dramatically changes the fit, splitting
the flux nearly equally between the \texttt{vapec} and \texttt{bbody} components. 
For both epochs, the best-fitting temperature of the blackbody component is $>2$\,keV---so high 
as to be unphysical for optically-thick emission on a white dwarf 
(SSS are not expected to exceed k$T \approx 0.2$\,keV; \citealt{2013ApJ...777..136W}).

Spectral fits to {\em Swift}/XRT observations of \nova{} (\S\,\ref{sec:xrtspectroscopy}) 
covering a wide time range (\S\,\ref{sec:swiftmonitoring}) do not require a second emission component 
and can be described as a single absorbed thermal plasma with temperature and absorbing column 
that gradually decrease with time. We take this as reassurance that there is no need to artificially 
introduce a second emission component for fitting the {\em NuSTAR} observations.


To constrain non-thermal X-rays, we fit the spectrum with an absorbed power-law model \texttt{vphabs*pow} 
(i.e., assuming that {\em all} X-ray emission is non-thermal---see the discussion in \S\,\ref{sec:nonthermalx}). 
The fit yields values of $\chi_{\rm red}^2=1.2$, ${\rm d.o.f.}=199$, $p=0.013$---slightly below 
our adopted significance level of $0.05$. The photon index for the best-fitting power-law model (Table~\ref{tab:nustarmodels}) 
is $\Gamma=3.9 \pm 0.1$ (\S\,\ref{sec:nonthermalx}). The associated absorbing column for the power law model is higher 
(by a factor of 1.5 for the XMM abundances model) than for the optically thin plasma model.

Finally, following \cite{2019ApJ...872...86N}, we test the possibility that
the {\em NuSTAR} emission is an (absorbed) combination of an  
optically-thin thermal plasma emission 
and non-thermal emission represented by a power-law (\texttt{vapec+pow}). 
From our absorbed plasma model fits, we see that a power-law component 
is not required to obtain an acceptable fit to the data; 
therefore, the model flux and photon index are not constrained if both 
the photon index and normalization factor are left free to vary. To circumvent
this, we consider three fixed values of the photon index: $\Gamma= 1.0$, $1.2$
and $2.0$. In all cases the contribution of the power-law
component is constrained at $\lesssim 2$\% of the thermal
component flux listed in Table~\ref{tab:nustarmodels}.

These tested photon index values are the ones expected for 
the low-energy tail of the GeV emission, as discussed in \S\,\ref{sec:nonthermalx}.
A different mechanism that may produce non-thermal X-ray emission in novae (that should operate independently of 
the process responsible for the GeV emission) is the Compton degradation of MeV $\gamma$-rays produced 
by radioactive decay \citep[see \S\,\ref{sec:xraynovaintro};][]{1992ApJ...394..217L,2010ApJ...723L..84S,2014ASPC..490..319H}. 
\cite{1998MNRAS.296..913G} predict flat or inverted (rising) continuum spectra below 100\,keV for the Comptonized photons 
in both CO and ONe novae (which differ by the set of parent radioactive decay lines). As with the low-energy tail of the GeV emission, 
the observed soft spectrum disfavours Comptonization of the radioactive lines as the source of X-ray emission from \nova{} 
in the {\em NuSTAR} band. 
\cite{2019ApJ...872...86N} argue that the Compton optical depth in a nova is not sufficient 
for Compton degradation to produce a detectable hard X-ray flux.
In summary, all the expected mechanisms behind non-thermal emission
should produce a hard spectrum, while in fact the observed spectrum is soft,
consistent with being thermal.

Finally, we construct an ``{\em XMM} abundances and fixed Galactic column''
model \texttt{constant*phabs*vphabs*vapec}
that includes a single emission component (\texttt{vapec}) and incorporates our knowledge of 
the elemental abundances (\S\,\ref{sec:ejectaabund}) and 
Galactic $N_\mathrm{H}$ (\texttt{phabs}; \S\,\ref{sec:extinction}).
We choose this as the preferred model 
(marked in boldface in Table~\ref{tab:nustarmodels}) 
for the {\em NuSTAR} spectra of \nova{}.

%
%

\subsection{{\em NuSTAR} variability search}
\label{sec:nustarlc}

We checked for the presence of variability within the two {\em NuSTAR}
observations that lasted 127 and 92\,ks wall time (total time including interruptions),
respectively (Table~\ref{tab:nustarlog}). The regular interruptions were caused by 
the Earth occultations of the source. 
For each of the two observations we generated source and background 
lightcurves with \texttt{nuproducts} using 5806\,s bin size 
(corresponding to one {\em NuSTAR} orbital revolution). The background
lightcurve was scaled and subtracted from the source lightcurve using
\texttt{lcmath}. We then performed the $\chi^2$ test to determine if the lightcurves 
are consistent with the null hypothesis that the source flux does not change 
during the observation given the errorbars. For a discussion of the $\chi^2$ test
in the context of variability search, see \cite{2010AJ....139.1269D,2017MNRAS.464..274S}. 
The test is sensitive to any kind of variability, both periodic and irregular.

We find that the null hypothesis cannot be rejected at the $3\sigma$ level, 
i.e., we found no significant variability within the individual {\em NuSTAR} observations. 
The r.m.s.\ scatter of the {\em NuSTAR} lightcurves is 0.008\,cts/s (18\%) and 0.007\,cts/s (6\%) for 
the first and the second epoch, respectively. If there is any low-level variability in the source during 
the times of our observations, the variability amplitude is lower than the above values. Our analysis probes the
variability time-scales from $\sim6$ to $\sim100$\,ks 
(variability related to orbital motion of the binary system might be
expected on these time-scales). 
Investigation of variability on a shorter time-scale is limited by S/N, while 
the upper bound on the detectable variability time-scale is set by the duration of our observations.
We leave the search for short time-scale periodic signals (that could be
associated with white dwarf rotation) outside the scope of this paper, 
as we do not expect the X-ray emitting nova shock to be physically tied 
to the white dwarf surface (for example - its magnetic pole).

\begin{figure}
\begin{center}
\includegraphics[width=0.48\textwidth,clip=true,trim=0cm 0cm 0cm 0cm,angle=0]{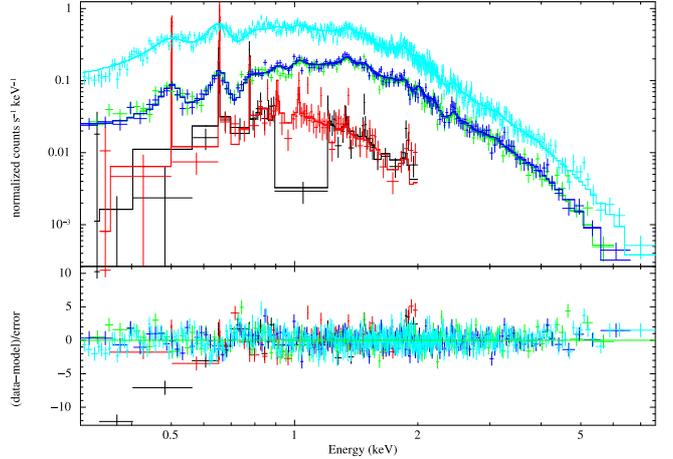}
\end{center}
\caption{{\em XMM-Newton} EPIC and (1st order) RGS spectra
of \nova{} obtained on
$t_0+275$. The colour coding is: 
black -- RGS1; 
red -- RGS2; 
green -- EPIC-MOS1; 
blue -- EPIC-MOS2; and  
cyan -- EPIC-pn. Solid lines represent the model described in Table~\ref{tab:xmm}.}
\label{fig:epicrgsspec}
\end{figure}

\begin{figure*}
\begin{center}
\includegraphics[height=1.0\textwidth,clip=true,trim=0cm 0cm 0.0cm 0cm,angle=270]{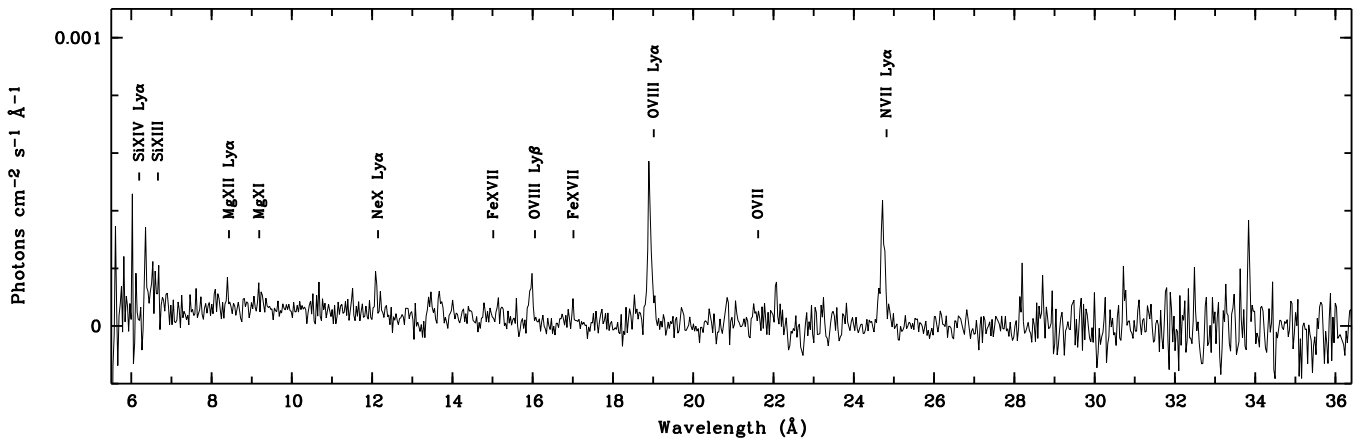}
\end{center}
\caption{{\em XMM-Newton} fluxed spectrum of \nova{} combining the 1st and 2nd order spectra from both RGS1 and RGS2.}
\label{fig:rgsspec}
\end{figure*}

\subsection{{\em XMM-Newton} spectroscopy}
\label{sec:xmm}

We requested an {\em XMM-Newton} target-of-opportunity observation to distinguish between 
the CNO-overabundant and Fe-deficient models that
both fit the {\em NuSTAR} spectra well, but differ in the predicted $N_\mathrm{H}$ value
by almost two orders of magnitude (\S\,\ref{sec:nustarspec}). A 51\,ks
observation (ObsID:0831790401) was conducted on 2018 December~16 ($t_0+275$\,d).
The observation was only partially ($\sim$\,10-20\% of the total effective exposure time) 
affected by high level of solar particles and all X-ray instruments collected useful
data (EPIC cameras: pn, MOS1, and MOS2; RGS cameras: RGS1 and RGS2).
The Optical Monitor had to be blocked due to the presence of the 
bright star HD\,92063 in its field of view, only 2\,$\arcmin$ from \nova{} 
(this was the star monitored by
BRITE and it also affected the {\em Swift}/XRT
observations; \S\,\ref{sec:discoveryinfo}, \S\,\ref{sec:swiftmonitoring}). The presence of HD\,92063
required the use of the thick optical blocking filter with the EPIC cameras. 
The data analysis was performed through the \texttt{XMM-Newton Science Analysis System (SAS)
v17.0.0}, using calibration files available in 2018 December. The EPIC data were grouped 
to have spectra with at least 25 counts per bin for each camera; 
for the RGS data the value was at least 5 counts per bin. 
The spectral fit, with \texttt{XSPEC~12.9.1m}, assumed the C-Statistic and Chi-Squared for the 
fit and test statistics, respectively.

The X-rays from \nova{} were
clearly detected with all EPIC and RGS instruments (Fig.~\ref{fig:epicrgsspec}). 
The average EPIC-pn count rate was $0.810 \pm 0.005$\,cts/s at the 0.3-8\,keV energy range, 
that corresponds to a total number of background corrected counts of 32679 for the
40350\,s of the exposure under low level of solar particle contamination. 
An X-ray source of this brightness should not produce any significant pile-up in any of the {\em XMM-Newton} instruments 
 and, in fact, there is no evidence of pile-up in the EPIC data from the \texttt{sas/epatplot} task.

The RGS spectra show a continuum and prominent emission lines, while there is no evidence of absorption 
lines (Fig.~\ref{fig:rgsspec}). Noticeably, among the emission lines there are the lines of
\ion{N}{vii} (K$\alpha$, analogous to the Lyman\,$\alpha$ line of hydrogen) and \ion{O}{viii}
(K$\alpha$ and K$\beta$), suggesting a high abundance of these elements.

The EPIC spectra (Fig.~\ref{fig:epicrgsspec}) do not show the 1\,keV ``bump'' expected from a number of 
Fe L-shell lines (analogous to the Balmer series), 
and there is no sign of the Fe\,K$\alpha$ feature -- suggesting sub-solar Fe abundance (\S\,\ref{sec:ejectaabund}). 
The C abundance could not be reliably constrained from the RGS data as the estimate
would rely on the \ion{C}{vi} Lyman\,$\alpha$ line at 33.7\,\AA\ (0.368\,keV) in the rest frame, that is 
located in the noisy part of the spectrum. However, 
for the solar C/N abundance ratio, the \ion{C}{vi} line 
should be stronger than the detected \ion{N}{vii} (24.8\,\AA, 0.500\,keV)
line \citep[e.g.][]{2001A&A...365L.329A} and should have been visible in our RGS
spectrum (Fig.~\ref{fig:rgsspec}).
The absence of the \ion{C}{vi} line, combined with the clear presence of the \ion{N}{vii} 
line imply that the C/N ratio is sub-solar.

The results of a joint fit to EPIC (0.3--8\,keV)
and RGS (0.65--2\,keV) spectra
of the absorbed thermal plasma model \texttt{phabs*vphabs*bvapec} with the
addition of Gaussian lines are presented in Table~\ref{tab:xmm}. 
The adopted abundance table was that of \cite{2009ARA&A..47..481A}. 
The Gaussian lines accounted for excesses that we associate with triplets of resonance (r), intercombination (i), forbidden
(f) lines of the He-like ions of \ion{Mg}{xi}, \ion{Ne}{ix}, \ion{O}{vii}, \ion{N}{vi}. 
The fitted line parameters are presented in Table~\ref{tab:xmmlines}.
The likely source of the discrepancy between what is predicted 
by the \texttt{bvapec} component and what is observed and associated to the triplets
is that the single-temperature equilibrium plasma is too crude an approximation of real
physical conditions in the nova ejecta that cannot fully describe the grating data.

We use a combination of solar abundances absorber (\texttt{phabs}) and
variable--abundances absorber (\texttt{vphabs}) 
to account for the Galactic and
intrinsic contributions to the total column density, respectively. 
We let the absorption column parameter  
for both absorbing components, expressed in the equivalent of hydrogen column ($N_\mathrm{H}$), to vary freely during the fit. 
The resulting Galactic $N_\mathrm{H}$
from X-rays is consistent at the 1\,$\sigma$ confidence level with the value estimated from
optical extinction (\S\,\ref{sec:extinction}), while the model shows that a non-negligible amount of material is also absorbing X-rays 
within the nova shell (Table~\ref{tab:xmm}).

The emission lines in Fig.~\ref{fig:rgsspec} appear blueshifted. 
This indicates that the bulk of the plasma responsible for the emission seen in X-rays is moving towards us.
Its radial velocity may be estimated from 
the redshift parameter of the \texttt{bvapec} component in the model.
A value of $-870\pm60$\,km\,s$^{-1}$ 
was derived from fitting the same data excluding the
energy ranges where the Gaussian lines had to be inserted in the final fit. 
Then this value was held fixed for the fit reported in Table~\ref{tab:xmm}.
Interestingly, had the optical lines been blueshifted by the same
velocity, that would have been easily noticeable in spectroscopic
observations by \cite{2020NatAs.tmp...79A}, but no such shift 
was observed\footnote{Systemic velocity of $-870\pm60$\,km\,s$^{-1}$ would be obvious 
in the optical spectra collected before the maximum light while the lines are not very broad 
(the dips of the P~Cygni profiles were at $\sim250$\,km\,s$^{-1}$). 
On $t_0 + 269$\,d, the FWHM of the Balmer lines was 
$\sim900$\,km\,s$^{-1}$ centered at $50 \pm 100$\,km\,s$^{-1}$ \citep{2020NatAs.tmp...79A}.}.
We do not have a conclusive explanation for this discrepancy, 
but it seems to be due to opacity and/or asymmetries in the ejecta. 
One possibility is that the ejecta are opaque to X-rays and we see only the approaching side of the expanding X-ray plasma, 
while it is fully transparent to photons in the optical by $t_0+275$. 
Alternatively, the X-ray emitting ejecta could be highly asymmetric, 
with the approaching part emitting much more than the receding part. 

A similar blueshift of X-ray emission lines was observed with {\em Chandra} 
by \cite{2008ApJ...673.1067N} in the red giant donor recurrent nova RS~Oph
and by \cite{2016ApJ...829....2P} in the classical GeV-bright nova V959\,Mon. 
Nelson et al. (2020, in prep.) confirm the blueshift with {\em Suzaku} spectroscopy
of V959\,Mon.
The opacity-based explanation of blueshifted emission lines in V959\,Mon suggested by \cite{2016ApJ...829....2P} is similar to that of 
the blueshifted {\it absorption} lines observed in the SSS spectra of other novae \citep{2012BASI...40..353N}.
\cite{2008ApJ...673.1067N} speculated that transient highly blueshifted \ion{C}{vi} 
and \ion{N}{vi} lines seen in RS~Oph may be associated with 
the asymmetric synchrotron-emitting jet observed with VLBI \citep{2008ApJ...688..559R}.

We use the \texttt{bvapec} model instead of \texttt{vapec} to account for 
the velocity- and thermal-broadened emission lines that become 
important when fitting the grating (RGS) spectra of \nova{}.
The line width for the final fit (Table~\ref{tab:xmm}) was fixed to the value 
derived from the preliminary fit that excluded data in the problematic regions where the Gaussian lines had to be added.
The line broadening derived from the preliminary fit was $\sigma = 378\pm72$\,km\,s$^{-1}$ (Gaussian sigma is the parameter of 
the \texttt{bvapec} model) corresponding to the full width at half maximum ${\rm FWHM} =  2 \sqrt{2 \ln{2}} \sigma = 890$\,km\,s$^{-1}$.

As with the {\em NuSTAR} spectra (\S\,\ref{sec:nustarspec}), we tried to add a
blackbody emission component to the optically thin thermal plasma model
described in Table~\ref{tab:xmm}. The resulting blackbody temperature
is unphysically high k$T\sim10$\,keV. We interpret it as the absence of any
SSS emission during the {\em XMM-Newton} observation. The optically thin
plasma plus blackbody model also does not fit {\em Swift}/XRT data taken
around this time (\S\,\ref{sec:xrtspectroscopy}).

\begin{table}
\caption{Parameters of the simultaneous \emph{XMM} (EPIC and RGS) spectra
fit with the model \texttt{constant*phabs*vphabs*bvapec} plus Gaussian lines
(Table~\ref{tab:xmmlines}).}
\label{tab:xmm}
\begin{tabular}{lccl}
\hline\hline
Parameter & Value & Comment \\
\hline                     
\texttt{phabs}       \\ 
$N_\mathrm{H}$ ($\times$10$^{21}$\,cm$^{-2}$)   & 2.4$^{+0.4}_{-0.3}$ & \\ 
\hline                     

\texttt{vphabs}      \\ $N_\mathrm{H}$ ($\times$10$^{21}$\,cm$^{-2}$)   & 0.12$^{+0.03}_{-0.03}$ & \\ 
\hline                     

\texttt{bvapec}      \\ $k$T (keV)                             & 1.07$^{+0.04}_{-0.01}$ & \\ 
                      redshift                                 & $-2.9\times10^{-3}$ & fixed \\ 
                      velocity (km\,s$^{-1}$)                  & 378  & fixed \\ 
\hline
                      N/N$_{\odot}$       & 345$^{+93}_{-70}$ & \\
                      O/O$_{\odot}$       & 29$^{+7}_{-5}$    & \\
                      Ne/Ne$_{\odot}$     & 2.2$^{+0.6}_{-0.5}$ & \\
                      Mg/Mg$_{\odot}$     & 0.6$^{+0.2}_{-0.1}$ & \\
                      Si/Si$_{\odot}$     & 1.1$^{+0.2}_{-0.2}$ & \\
                      Fe/Fe$_{\odot}$     & $<$0.1 & \\
\hline                     
$\chi^{2}_{\nu}$ & 1.15 & \\
d.o.f.  & 1837 & \\ 
\hline
\end{tabular}
\end{table}


\begin{table}
\caption{Gaussian lines added to the \texttt{bvapec} model}
\label{tab:xmmlines}
\begin{tabular}{ccccc}
\hline\hline
Lines & & Energy & Wavelength & Line flux  \\
      & & (keV)  &  (\AA)   & ~photons~cm$^{-2}$~s$^{-1}$ \\
\hline
\ion{Mg}{xi} & r &  9.169 & 1.356 & $5.2\pm2.1$  \\
             & i &  9.235 & 1.346 & $0$  \\
             & f &  9.314 & 1.335 & $3.5^{+2.1}_{-2.0}$  \\
\ion{Ne}{ix} & r & 13.447 & 0.925 & $0$  \\
             & i & 13.551 & 0.918 & $<5.0$  \\
             & f & 13.698 & 0.908 & $14.0^{+2.9}_{-2.8}$  \\
\ion{O}{vii} & r & 21.602 & 0.576 & $8.4\pm4.0$  \\
             & i & 21.802 & 0.570 & $5.0^{+3.9}_{-3.5}$  \\
             & f & 22.097 & 0.563 & $16.5^{+4.6}_{-4.5}$  \\
\ion{N}{vi}  & r & 28.792 & 0.432 & $13.9^{+4.8}_{-4.7}$  \\
             & i & 29.074 & 0.428 & $0$  \\
             & f & 29.531 & 0.421 & $<9.5$  \\
\hline
\end{tabular}
\begin{flushleft}
\end{flushleft}
\end{table}

\subsection{{\em Swift}/XRT monitoring}
\label{sec:swiftmonitoring}

{\em Swift} observed \nova{} on 45 epochs between 2018 March~21 ($t_0+5.4$\,d) 
and 2019 June~8 ($t_0+449$\,d). The first three observations on 2018 March~21, April~22, and May~11 
resulted in non-detections. The less-sensitive Windowed Timing mode had to be used in the first two
observations to reduce optical loading while the nova was still optically bright.
During the third observation, the XRT was automatically switching between the Windowed Timing (17\,s exposure)
and Photon Counting (277\,s exposure) modes.
\nova{} was clearly detected in the 42 following observations
(starting from 2018 May~17, $t_0+63$\,d), all performed in the Photon Counting mode
with a typical exposure time of 1.5\,ks (46.7\,ks total exposure).
The observations on 2018 September~30 and 2018 November~18
have low signal-to-noise, as the XRT image of the source was crossed by a bad CCD column reducing the number of detected photons.

We use the standard circular source extraction region with a radius of
20\,pix centred at the position derived from Swift/UVOT astrometry (\S\,\ref{sec:uvot}).
For the background we use an annulus centred on the source position with an inner radius 
of 77\,pix and outer radius of 101\,pix. This non-standard background extraction region was chosen 
to avoid the two nearby X-ray sources (clearly visible in the stacked image) and the cluster 
of optical photons from the nearby bright ($V=5.09$) star HD\,92063.
We use only grade~0 events in the analysis in order to minimize optical loading.

\begin{figure}
\begin{center}
\includegraphics[width=0.48\textwidth,clip=true,trim=0cm 0cm 0cm 1cm]{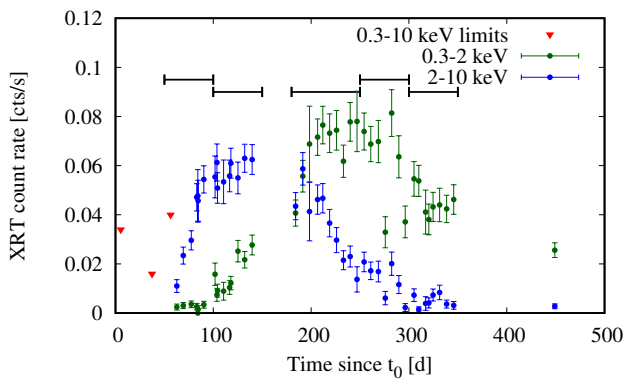}
\end{center}
\caption{{\em Swift}/XRT lightcurve of \nova{}. 
The red triangles mark the 0.3--10\,keV upper limits derived from the Windowed Timing mode observations.
The horizontal bars indicate the width of time bins used for the spectral analysis (Table~\ref{tab:swiftmodels}).}
\label{fig:swiftlc}
\end{figure}

Fig.~\ref{fig:swiftlc} presents the {\em Swift}/XRT lightcurve of \nova{} in the soft (0.3--2\,keV) and hard (2--10\,keV) bands. 
The hard flux is steeply rising following the initial detection on 2018 May~11
($t_0+63$\,d), reaches a plateau around 2018 June~14 ($t_0+90$), and then declines after 2018 September~23 ($t_0+191$\,d). 
The soft flux gradually rises from 2018 June~25 ($t_0+102$\,d) 
until 2018 October~8 ($t_0+207$\,d), then---after a standstill---it starts to decline on 2018 December~9 ($t_0+269$\,d).
The peak full band (0.3-10\,keV) count rate was reached on 2018 October~14 ($t_0 + 212$\,d) at $0.13 \pm 0.01$\,cts~s$^{-1}$, and
is sufficiently low that no pile-up correction is required.

\begin{table*}
        \centering
        \caption{Parameters of the {\em Swift}/XRT spectral models using {\em XMM}-derived abundances}
        \label{tab:swiftmodels}
        \begin{tabular}{c@{~~~~~~}c@{~~}c@{~~}c@{~~}c@{~~}c@{~~}c} 
                \hline
Epoch & \texttt{vphabs} $N_\mathrm{H}$                     & k$T$        &  0.3--10.0\,keV Flux              & $\chi_{\rm red}^2$ & d.o.f. & $p$  \\ 
(days)& ($10^{22}$\,cm$^{-2}$) & (keV)       &  $\log_{10}$(erg\,cm$^{-2}$\,s$^{-1}$)  &                    &        &      \\
                \hline
 \multicolumn{7}{c}{Model \texttt{phabs*vphabs*vapec}} \\
050--100  &  $ 0.61 \pm  0.12 $  &  $68 \pm  452$  &  $ -11.155 \pm 0.145 $ & 0.64 &   12 & 0.81 \\
100--150  &  $ 0.160 \pm  0.015 $  &  $ 6.2 \pm  1.4 $  &  $ -11.148 \pm 0.028 $ & 0.99 &   32 & 0.48 \\
180--250  &  $ 0.019 \pm  0.002 $  &  $ 2.18 \pm  0.24 $  &  $ -11.288 \pm 0.018 $ & 1.06 &   40 & 0.37 \\
250--300  &  $ 0.017 \pm  0.003 $  &  $ 0.98 \pm  0.12 $  &  $ -11.705 \pm 0.021 $ & 1.11 &   18 & 0.33 \\
300--350  &  $ 0.009 \pm  0.003 $  &  $ 0.74 \pm  0.12 $  &  $ -11.916 \pm 0.023 $ & 1.00 &   14 & 0.45 \\
 \multicolumn{7}{c}{Model \texttt{phabs*vphabs*bbody}} \\
050--100  &  $ 0.269 \pm  0.080 $  &  $ 2.19 \pm  0.42 $  &  $ -11.140 \pm 0.032 $ & 0.42 &   12 & 0.96 \\
100--150  &  $ 0.064 \pm  0.012 $  &  $ 1.15 \pm  0.06 $  &  $ -11.198 \pm 0.022 $ & 0.82 &   32 & 0.76 \\
180--250  &  $ 0.000 \pm  0.001 $  &  $ 0.61 \pm  0.02 $  &  $ -11.341 \pm 0.015 $ & 1.19 &   40 & 0.19 \\
250--300  &  $ 0.000 \pm  0.003 $  &  $ 0.39 \pm  0.02 $  &  $ -11.751 \pm 0.020 $ & 1.83 &   18 & 0.02 \\
300--350  &  $ 0.000 \pm  0.001 $  &  $ 0.27 \pm  0.01 $  &  $ -12.006 \pm 0.021 $ & 3.33 &   14 & 0.00 \\
                \hline
        \end{tabular}
\begin{flushleft}
{\bf Column designation:}
Col.~1~-- time since outburst;
Col.~2~-- equivalent hydrogen column density;
Col.~3~-- plasma or blackbody temperature;
Col.~4~-- absorbed 0.3--10.0\,keV flux;
Col.~5~-- reduced $\chi^2$;
Col.~6~-- number of degrees of freedom;
Col.~7~-- Null hypothesis probability.
\end{flushleft}
\end{table*}

\subsection{{\em Swift}/XRT spectroscopy}
\label{sec:xrtspectroscopy}

\begin{figure}
\begin{center}
\includegraphics[width=0.48\textwidth,clip=true,trim=0cm 0.5cm 0.5cm 0.5cm,angle=0]{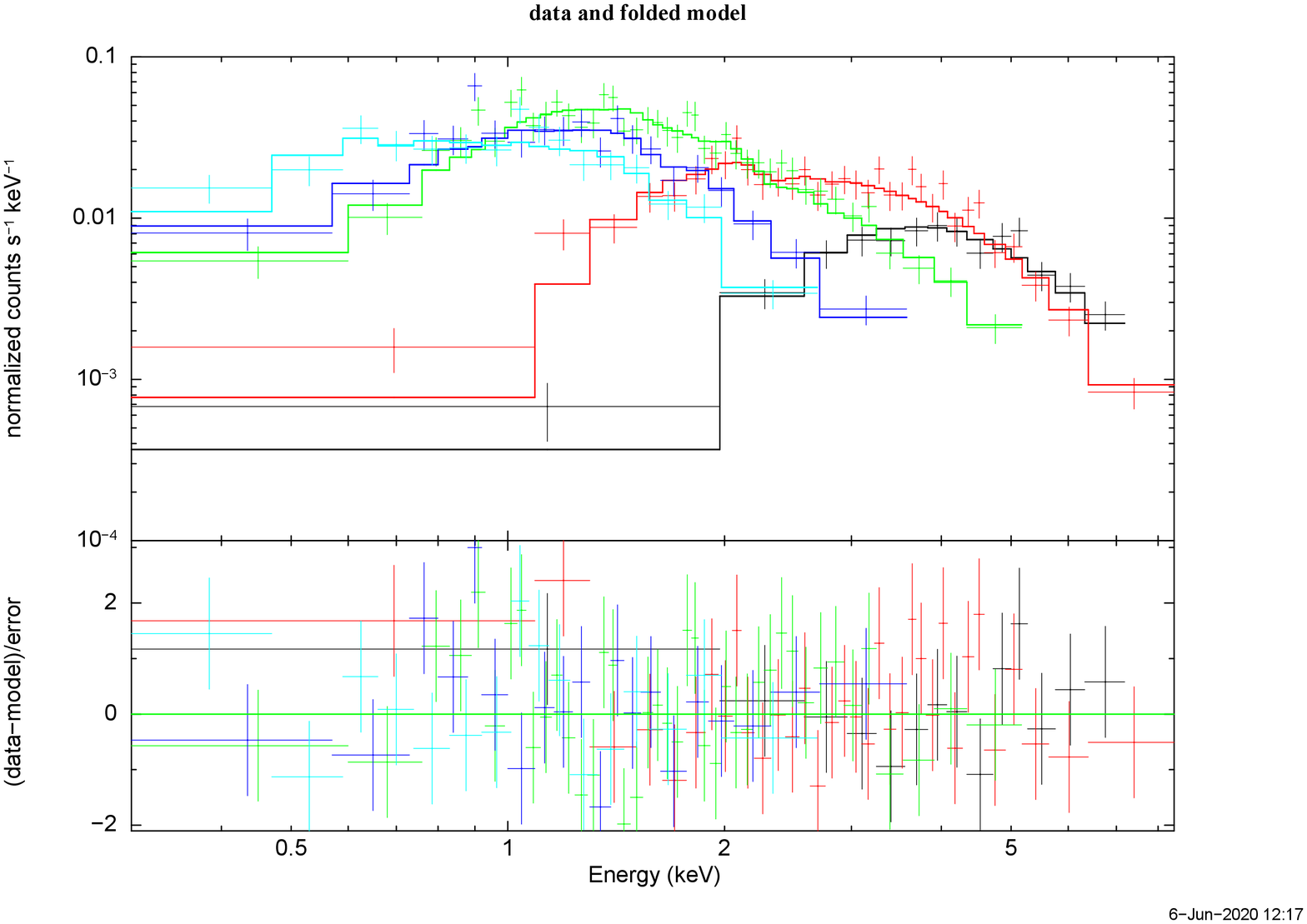}
\end{center}
\caption{{\em Swift}/XRT spectra of \nova{} obtained 
50--100 (black), 
100--150 (red), 
180--250 (green), 
250--300 (blue), 
300--350 (cyan)
days after $t_0$.
Solid lines represent \texttt{phabs*vphabs*vapec} models described in Table~\ref{tab:swiftmodels}.}
\label{fig:xrtspecspecev}
\end{figure}

To follow the spectral evolution of \nova{}, we construct five spectra by combining 
{\em Swift}/XRT observations taken within $\sim50$\,day intervals marked in Fig.~\ref{fig:swiftlc} 
(see Table~\ref{tab:swiftmodels}). 
The spectra are presented in Fig.~\ref{fig:xrtspecspecev}.
We binned individual {\em Swift}/XRT observations to increase 
the photon statistics \citep[e.g.][]{2012ApJ...748...43N}, but we note the X-ray spectrum is changing within each bin. 
This may degrade the quality of the fits reported in Table~\ref{tab:swiftmodels}. 
The bin width of $\sim50$\,days was chosen as a compromise between the photon statistics and the rate of spectral
changes. 

We fit the data with absorbed thermal plasma models (\texttt{phabs*vphabs*vapec}) and absorbed blackbody models
(\texttt{phabs*vphabs*bbody}), 
fixing the elemental abundances in the \texttt{vphabs} and \texttt{vapec}
components to the values derived from our {\em XMM-Newton} spectroscopy (\S\ref{sec:xmm}; Table~\ref{tab:xmm})
and the Galactic absorption \texttt{phabs} (having the solar abundances) to the expected value (\S\ref{sec:extinction}). 
The intrinsic absorbing column ($N_\mathrm{H}$) and plasma/blackbody temperature (k$T$), as well as the emitting component flux, 
are left as free parameters. 
The column densities reported in Table~\ref{tab:nustarmodels} and Table~\ref{tab:swiftmodels} refer only to 
the variable intrinsic \texttt{vphabs} absorption component.
The absorbed blackbody fit suggests no intrinsic absorption 
after $t_0+150$\,d, but fails to provide a good fit for the last two Swift spectra (day 250--300 and 300--350). 
In addition, the blackbody temperatures are unphysically high for an SSS \citep{2013ApJ...777..136W}. 
The absorbed optically-thin plasma fits all five spectra well, 
and provide a physically appropriate model for the X-ray emission. 
The optically thin model is also supported by the {\em NuSTAR} (\S\ref{sec:nustarspec}) 
and {\em XMM-Newton} (\S\ref{sec:xmm}) observations. Therefore, we assume that the emission 
was dominated by the optically-thin component at all times.

\subsection{{\em Swift}/UVOT lightcurve and astrometry}
\label{sec:uvot}

Most of the optical/UV photometry collected with UVOT during 
the {\em Swift} monitoring was not useful due to high coincidence losses on the bright source. 
Only starting from $t_0+212$\,d does the source become sufficiently faint to perform photometry 
in the uvm2 band. The UV lightcurve shows a smooth decline from uvm2=10.44\,mag on 2018 October~14 ($t_0+212$\,d) to uvm2=11.78\,mag on 2019 June~8 ($t_0+449$\,d). The photometric error is dominated by the uncertainty in the coincidence-loss correction and is expected to be at the level of a few percent.
We used these uvm2 images obtained during the decline phase to measure the astrometric position of \nova{} 
relative to UCAC3 stars within the UVOT field of view \citep{2010AJ....139.2184Z}. 
We measured the nova position with the uncertainty of $\sim$0.1\,$\arcsec$ 
(estimated from the scatter of measurements from multiple images):
\begin{verbatim}
10:36:15.42 -59:35:54.0 J2000.
\end{verbatim}

\subsection{{\em Fermi}/LAT monitoring}
\label{sec:latobs}

\cite{2020NatAs.tmp...79A} performed detailed analysis of {\em Fermi}/LAT observations of \nova{}, 
establishing it as the brightest $\gamma$-ray emitting nova observed to date, detected during 2018 April~8--30. 
Unfortunately, the observations were cut short by the failure of the solar panel drive on 2018 March~16 (around $t_0$) 
that sent the {\em Fermi} spacecraft to ``safe mode'' with the scientific instruments powered off. 
The LAT observations resumed on 2018 April~8 ($t_0+23$\,d), but were interrupted again for the period 2018 May~1--13 
(45--58 days after $t_0$)  by a {\em Fermi} pointing pattern (needed to keep the stuck solar panel illuminated) 
that was unfavourable for observations of \nova{}. 

Using the power law with exponential cut-off model for the $\gamma$-ray spectrum presented in \cite{2020NatAs.tmp...79A} and
restricting the LAT exposure to the exact time range of the first {\em NuSTAR} observation (Table~\ref{tab:nustarlog}), 
we derive a significant $\gamma$-ray detection of \nova{} (test statistic $TS=283$, or $\sim 17\sigma$ detection;
\citealt{1996ApJ...461..396M}). Its 100\,MeV--300\,GeV photon flux is $(1.17 \pm 0.11) \times 10^{-6}$\,photons\,cm$^{-2}$\,s$^{-1}$, 
and $\nu F_\nu = (1.00 \pm 0.10) \times 10^{-10}$\,erg cm$^{-2}$\,s$^{-1}$ at photon energy h$\nu = 100$\,MeV, where h is the Planck constant.

The second {\em NuSTAR} observation (Table~\ref{tab:nustarlog}) was performed a day before 
LAT resumed observations of the \nova{} region. So, instead of using strictly simultaneous data, 
we use the LAT data collected right after the coverage gap, between 2018 May~13 17:01:00 and
May~15 04:16\,UT (same exposure time as the duration of the
first {\em NuSTAR} observation including interruptions). 
The LAT observations resulted in a non-detection of \nova{} ($TS=0$), 
with an upper limit on the 100\,MeV--300\,GeV photon flux of 
$<1.64 \times 10^{-7}$\,photons\,cm$^{-2}$\,s$^{-1}$ and 
$\nu F_\nu < 1.41 \times 10^{-11}$\,erg cm$^{-2}$\,s$^{-1}$ at h$\nu = 100$\,MeV.

%
%
%
%

\section{Discussion}
\label{sec:discussion}

\subsection{Ejecta abundances}
\label{sec:ejectaabund}

It has long been recognized that nova ejecta are often enriched in heavy elements, 
compared to the composition of matter accreted from the donor star 
\citep[e.g.,][]{1986ApJ...308..721T,1998PASP..110....3G,2012ApJ...755...37H}. 
This chemical enrichment is attributed to mixing between the accreted material 
and the white dwarf itself \citep{2008clno.book.....S}.
Computer simulations have demonstrated that this mixing probably occurs at the onset 
of the thermonuclear runaway due to Kelvin-Helmholtz instabilities 
\citep{2011Natur.478..490C,2016A&A...595A..28C,2018A&A...619A.121C}.
Thermonuclear burning in the nova event proceeds through the CNO cycle, 
and may change the relative abundances of C, N and O, 
but will not increase the overall abundance of 
the CNO elements \citep{1972ApJ...176..169S,1986ApJ...308..721T}. 
The super-solar N/C ratio found in \nova{} (\S~\ref{sec:xmm}) 
demonstrates that the plasma emitting the soft X-ray lines 
has undergone non-equilibrium CNO burning. The plasma was ejected  
while in the $^{14}$N bottleneck \citep[e.g.][]{2004A&A...420..625I}.
Optical spectroscopy often finds in novae an overabundance of nitrogen by two
orders of magnitude and oxygen by one order of magnitude  
compared to solar values, respectively \citep{1998PASP..110....3G,2002AstL...28..100A,2013ApJ...762..105D}.
However, different authors sometimes report quite different abundances for
the same nova \citep[for example, compare the abundances for V1974\,Cyg
reported by ][]{1996AJ....111..869A,1996ApJ...469..854H,1997AstL...23..713A,2005ApJ...624..914V};
the same is true for the few X-ray derived abundances 
\citep[e.g.][]{2010ApJ...717..363R,2010PhDT.........6N}.

The chemical composition of the nova ejecta is strongly affected
by the composition of the white dwarf, because 
we do not generally expect the material accreted from the donor star 
to have super-solar CNO abundances or to be hydrogen-deficient.
Therefore, the white dwarf material is being ablated during a nova eruption
\citep[e.g.][]{2018ApJ...860..110S} and we can draw conclusions about 
the composition of the white dwarf by observing nova ejecta.

Depending on its zero-age main sequence mass and mass transfer due to binary evolution, 
the white dwarf hosting nova eruptions may have either a CO or ONe composition.
Our {\em XMM-Newton} spectroscopy of \nova{} implies CNO abundances that are 
a factor of $\sim$100 super-solar, but near-solar abundance of Ne (Table \ref{tab:xmm}). 
This suggests that the \nova{} host is a CO white dwarf. The dust formation episode exhibited 
50--100 days after outburst is consistent with a CO white dwarf \citep{2020NatAs.tmp...79A}, 
as dust formation is more common in nova ejecta enriched in CO than in ONe \citep{Evans_Rawlings_2008}.
The CO composition is consistent with a low mass of the white dwarf hinted at
by the non-detection of the super-soft emission (\S\,\ref{sec:missingsss}). 
The {\em XMM-Newton} spectroscopy also suggests sub-solar Fe abundance (Table \ref{tab:xmm}).

\subsection{The absence of the Fe~K$\alpha$ feature}

We found no evidence of the Fe~K$\alpha$ emission in \nova{}. This is not surprising given that the absence of Fe~K$\alpha$ emission in the X-ray spectra of novae is a long-standing puzzle.
At 6.7\,keV, this feature probes shock-heated plasma.
The two clear examples of novae with no Fe~K$\alpha$ emission are 
V382\,Vel (\S~4.3 in \citealt{2001ApJ...551.1024M}; 
also no Fe~L-shell emission found by \citealt{2005MNRAS.364.1015N}) 
and V959\,Mon (Nelson et al.\ 2020, in~prep.). 
The weakness of the Fe~K$\alpha$ emission may result from
a low abundance of iron in nova ejecta, 
a high abundance of CNO that would enhance the continuum making the Fe~K$\alpha$ line relatively weaker,
or from a non-equilibrium ionization state of the emitting plasma. 
The Fe\,II optical spectroscopic type reported for V382\,Vel by \cite{1999IAUC.7193....1D} and \cite{1999IAUC.7185....2S} 
led \cite{2001ApJ...551.1024M} to the conclusion that 
a non-equilibrium ionization state is responsible for the weakness of the iron line. 
V959\,Mon was a neon nova with sub-solar iron abundance \citep{2013A&A...553A.123S}.
Novae occurring in symbiotic star systems that include a giant (rather than main sequence) companion to the white dwarf 
show strong Fe~K$\alpha$ emission (RS\,Oph and V745\,Sco;
\citealt{2008ApJ...673.1067N,2009AJ....137.3414N,2015MNRAS.448L..35O,2019MNRAS.490.3691D}). 
Such emission was also observed in V2491\,Cyg by
\cite{2009ApJ...697L..54T}, \cite{2011PASJ...63S.729T}, 
but the nature of the companion star in this system is uncertain. 
In many cases, the composition of the emitting plasma
(rather than its ionization state) determines the strength of the Fe~K$\alpha$ emission.

\subsection{Ejecta mass}
\label{sec:ejectamass}

Observational estimates of how much material is ejected by novae provide a fundamental test of nova models.
Ejecta masses, when combined with the Galactic nova rate, constrain the contribution of novae to the chemical evolution of the Galaxy,
especially isotopes like $^7$Li, $^{13}$C, $^{15}$N, $^{17}$O
\citep{2013ApJ...762..105D,2016PASJ...68...39L, 2016MNRAS.463L.117M}. The abundances of these isotopes 
allow laboratory identification of nova dust grains in meteorites \citep{2018ApJ...855...76I}.

We can use $N_\mathrm{H}$ values derived from the X-ray spectral fitting 
as a function of time after outburst to estimate the ejecta mass. 
We assume that the source of X-rays is embedded deep into the ejecta, shining through most of it (\S~\ref{sec:shocklocation}).
The ejecta are modelled as a ``Hubble flow'', where the mass is expelled in 
a single impulse and is uniformly distributed over a range of velocities spanning 
from $v_{\rm min}$ to $v_{\rm max}$. 
This corresponds to the mass density $\propto r^{-2}$
\citep[e.g.,][]{Seaquist_Bode_2008}.
For $v_{\rm max}$, we take the maximum expansion velocity measured from the wavelength difference 
between the absorption dip and the emission peak of the P~Cygni profiles of Balmer lines,
$2500 \pm 100$\,km\,s$^{-1}$ (measured around $t_0+24$\,d;
\citealt{2020NatAs.tmp...79A}, Harvey et al.\ 2020, in prep.). 
In the Hubble flow model the slower moving ejecta will dominate the $N_\mathrm{H}$ at later epochs, 
when the faster moving ejecta have dispersed; the total ejecta mass in this model critically depends 
on the choice of $v_{\rm min}$.
Following \cite{2014ApJ...788..130C} we assume $v_{\rm min} = 0.2 v_{\rm max}$ which is in the middle of 
the range of values reported in the literature from modelling the multi-frequency radio
lightcurves of novae \citep{Seaquist_Bode_2008, 2016MNRAS.457..887W, 2016MNRAS.460.2687W, 2018ApJ...852..108F}; 
see Appendix~\ref{sec:vminvmaxliterature}.

We apply the Hubble flow model to the $N_\mathrm{H}$ evolution derived from the {\em NuSTAR} spectra fitted with the 
preferred model (Fig.~\ref{fig:nhtime}). 
Assuming that the ejecta began expanding at $t_0$, there is no acceptable fit to the $N_\mathrm{H}$ evolution 
(see dashed line in Fig.~\ref{fig:nhtime}). 
A good fit can be achieved if we assume that the mass was ejected at $t_0+24$\,d (solid curve in Fig. \ref{fig:nhtime}). 
Other novae have shown evidence for ejection of mass delayed weeks to months after the start of outburst: 
V2362\,Cyg \citep{2008A&A...479L..51K,2008AJ....136.1815L,2010PASJ...62.1103A},
T\,Pyx \citep{2014ApJ...785...78N,2014ApJ...788..130C} and V959\,Mon (\citealt{2014Natur.514..339C,Linford_etal15}, 
Nelson et al.\ 2020 in prep.).
In this model, the ejecta mass is $2.8\times10^{-5}$\,M$_\odot$ assuming spherical symmetry and the range of expansion velocities 
between $v_{\rm max} = 2500$\,km\,s$^{-1}$ and $v_{\rm min} = 0.2 v_{\rm max}$. The ejecta mass estimate strongly depends on these assumptions. 
Setting $v_{\rm max} = 600$\,km\,s$^{-1}$ (the slow component observed by \citealt{2020NatAs.tmp...79A}) would decrease the ejecta mass estimate 
by a factor of 20. Setting $v_{\rm min} = 0.1 v_{\rm max}$ would increase the mass estimate by a factor of four.
In the above ejecta mass calculation we counted only the hydrogen atoms.
The hydrogen mass should be multiplied by a factor of 1.90 for the derived
nova abundances (Table~\ref{tab:xmm}; the factor would be 1.36 for the solar
abundances of \citealt{2009ARA&A..47..481A}).

Taking into account the chemical composition of the ejecta and setting 
$v_{\rm min} \ll 0.2 v_{\rm max}$ is needed to reconcile the above X-ray
absorption-based ejecta mass estimate with the lower limit of
$2\times10^{-4}$\,M$_\odot$ derived from radio observations of \nova{} by
\cite{2020NatAs.tmp...79A} and the optical spectroscopy-based estimate 
of $6\times10^{-4}$\,M$_\odot$ by \cite{2020MNRAS.495.2075P}. Alternatively,
the discrepancy may be attributed to the X-ray emitting region being located 
above a considerable part of the ejecta (\S~\ref{sec:shocklocation}) or
ejecta being asymmetric (our line of sight may have less-than-average amount 
of X-ray absorbing material). Ejecta mass estimates from the radio lightcurve 
and optical spectroscopy are also subject to their own model assumptions
(such as clumpiness and temperature distribution).

Our conclusion about the delayed ejection depends on the assumption of 
a one-time ejection with a range of velocities. Other scenarios
may not require the delayed ejection and imply substantially different ejecta mass
given the same observed $N_\mathrm{H}$.
The mass may be continuously ejected over an extended period of time. 
The ejecta may experience (a period of) continuous acceleration 
(e.g. slow circumbinary material gradually pushed away by the fast white dwarf wind).
The ejecta may be asymmetric. The observation that $N_\mathrm{H}$ values
measured with {\em Swift} and {\em XMM-Newton} up to $t_0+350$\,d generally
follow the trend predicted by the delayed one-time ejection model fitting
{\em NuSTAR} observations on $t_0+36$\,d and $t_0+57$\,d (Fig.~\ref{fig:nhtime})
suggests, that this simple model may be a reasonable approximation of the actual ejection.

            

\subsection{The $L_X/L_\gamma$ ratio}
\label{sec:lxlg}

The ratio between the X-ray luminosity, $L_X$, and $\gamma$-ray luminosity, $L_\gamma$, 
provides insights into the physics of shocks in novae.

\subsubsection{Non-thermal X-rays}
\label{sec:nonthermalx}

\cite{2018ApJ...852...62V} discuss the possibility of non-thermal X-rays in
novae, predicting the spectral slope $\nu F_{\nu} \propto \nu^{0.8}$ ($\Gamma = 1.2$; \S~\ref{sec:thispaper})
in the hard X-ray band ($\gtrsim10$\,keV).
The observed {\em NuSTAR} spectrum of \nova{} is very soft, having $\Gamma > 2$ when fit with a single power-law
(\S\,\ref{sec:nustarspec}). This supports the interpretation that the observed X-ray emission is thermal, 
rather than non-thermal.
Another key prediction of \cite{2018ApJ...852...62V} is the existence of a lower limit
on the ratio of non-thermal X-ray to $\gamma$-ray fluxes. In $\nu F_{\nu}$ units
the limits are $L_X/L_\gamma > 10^{-3}$ for a leptonic origin of $\gamma$-rays, 
and $L_X/L_\gamma > 10^{-4}$ for the hadronic model.

Table~\ref{tab:lxlgammaratiotable} summarizes the available $L_{\rm 20\,keV}/L_{\rm 100\,MeV}$ measurements.
{\em NuSTAR} observations of novae V339\,Del and V5668\,Sgr, carried out during the GeV-bright
phase, yielded non-detections in the hard X-ray band (Mukai~et~al. in prep.). \cite{2018ApJ...852...62V} used 
these upper limits on the $L_X/L_\gamma$ ratio 
to suggest that the hadronic rather than leptonic mechanism is responsible for the $\gamma$-ray emission of novae.
\cite{2019ApJ...872...86N} observed the nova V5855\,Sgr with {\em NuSTAR} twelve days after eruption, 
while it was still detected in $\gamma$-rays by {\em Fermi}/LAT. V5855\,Sgr is the first nova in which $>10$\,keV X-rays 
and $\gamma$-rays were detected simultaneously. 
The {\em NuSTAR} spectrum of V5855\,Sgr is mostly featureless and can be fit by either an absorbed bremsstrahlung 
(all emission in {\em NuSTAR} band is thermal) or an absorbed power law (all emission is nonthermal) model. 
\cite{2019ApJ...872...86N} prefer the thermal emission model by 
invoking the spectral slope argument (that low-energy
tail of GeV emission should result in a hard {\em NuSTAR} spectrum 
while the best-fitting power law has $\Gamma = 3.6^{+1.3}_{-1.0}$).
Therefore, the $L_{\rm 20\,keV}/L_{\rm 100\,MeV}$ ratio 
measured by \cite{2019ApJ...872...86N} is considered by the authors an upper limit
on any non-thermal emission (Table~\ref{tab:lxlgammaratiotable}). This upper limit in V5855~Sgr is consistent with both the
leptonic and hadronic scenarios \citep{2018ApJ...852...62V}.

\begin{figure}
\begin{center}
\includegraphics[width=0.48\textwidth,clip=true,trim=0cm 0cm 0cm 0cm]{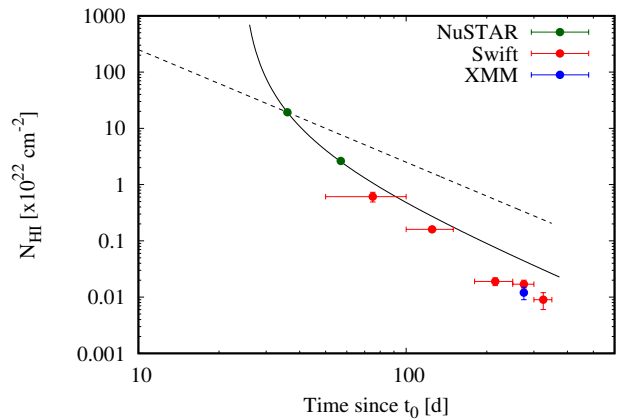}
\end{center}
\caption{The evolution of the absorbing column intrinsic to \nova{} as a function of time.
The solid curve represents the delayed ejection model fitting {\em NuSTAR} observations.
The dashed line represents the model with prompt ejection at $t_0$, that
predicts the same column density on $t_0+36$ as the delayed ejection model.}
\label{fig:nhtime}
\end{figure}

\begin{table}
\caption{X-ray to $\gamma$-ray monochromatic flux ratio in $\nu F_{\nu}$ units}
\label{tab:lxlgammaratiotable}
\begin{tabular}{ccc}
\hline\hline
Nova       & $L_{\rm 20\,keV}/L_{\rm 100\,MeV}$ & Reference \\
\hline
V339\,Del  & $< 4.0 \times 10^{-3}$ & \cite{2018ApJ...852...62V} \\
V5668\,Sgr & $< 1.7 \times 10^{-3}$ & \cite{2018ApJ...852...62V} \\
V5855\,Sgr & 0.01                   & \cite{2019ApJ...872...86N} \\
\nova{}    & 0.02                   & this work \\
\hline
\end{tabular}
\begin{flushleft}
\end{flushleft}
\end{table}

\subsubsection{Thermal X-rays}
\label{sec:thermalx}

Thermonuclear reactions heat the white dwarf atmosphere, but only to temperatures $<0.2$\,keV \citep{2013ApJ...777..136W}, 
which is identified with super-soft X-ray emission (\S\,\ref{sec:xraynovaintro}). 
Thermal emission of novae with k$T \gtrsim 0.5$\,keV is attributed to optically-thin shock-heated
plasma. The Rankine--Hugoniot conditions for a strong shock propagating in monoatomic gas
(with polytropic exponent 5/3) relate the post-shock temperature ($T_{\rm shock}$) to the shock velocity ($v_{\rm shock}$):
\begin{equation}
k T_{\rm shock} = \frac{3}{16} \mu m_p v_{\rm shock}^2
\label{eq:ktshock}
\end{equation}
(Equation~(30) of
\citealt{2014MNRAS.442..713M}), where $m_p$ is the proton mass, $k$ is the Boltzmann constant, $\mu$ is the mean molecular weight. 
%
For the fully ionized gas with the solar abundances as derived 
by \cite{2009ARA&A..47..481A}, $\mu=0.60$. 
Meanwhile, $\mu=0.74$ for the nova abundances from Table~\ref{tab:xmm}.
From (\ref{eq:ktshock}) we conclude that $v_{\rm shock}$ decreased from
2400 to 1700\,km\,s$^{-1}$ between the two {\em NuSTAR} epochs (Table~\ref{tab:nustarmodels}).
The shock velocity of 2400\,km\,s$^{-1}$ observed on day 36 is consistent with the velocity of the fastest wind, 
which starts around day 23 and remains visible in the optical line profile until at least day 35 \citep{2020NatAs.tmp...79A}.
The hypothetical second emitting component with k$T = 0.6$\,keV considered in the two temperature plasma model 
described in \S\,\ref{sec:nustarspec} would correspond to $v_{\rm shock} = 640$\,km\,s$^{-1}$.
We point out that $v_{\rm shock}$ is different from the ejecta expansion velocity, 
unless the ejecta slam into pre-existing low-velocity material. 
If the shock is formed at the interface between the slow and fast wind, 
$v_{\rm shock}$ would correspond to the velocity difference between the two components.

Radiative shocks should produce thermal X-ray emission that is at least 1--2 orders of magnitude brighter (in $\nu F_{\nu}$ units) 
than the GeV emission of the non-thermal particles \citep{2015MNRAS.450.2739M}. 
This is in sharp contrast with our {\em NuSTAR} observations, which show that the X-ray emission is almost two orders of magnitude fainter 
than the {\em Fermi}/LAT emission: $L_{\rm 20\,keV}/L_{\rm 100\,MeV} = 0.02$ (\S\,\ref{sec:nonthermalx}).
A similar discrepancy is reported by \cite{2019ApJ...872...86N} for V5855\,Sgr 
and even stronger ones are implied by the non-detections of V339\,Del and V5668\,Sgr
(Table~\ref{tab:lxlgammaratiotable}; Mukai~et~al. in prep.).
Thermal X-rays from the radiative shock can be suppressed thanks to a corrugated shock front
geometry \citep{2018MNRAS.479..687S} or by redistributing
emission thanks to Compton scattering in a highly non-spherical nova ejecta \citep{2019ApJ...872...86N}.
Each of the two effects can suppress the X-ray emission by an order of
magnitude, which is still not sufficient to account for the 3--4 orders of
magnitude difference between the predictions \citep{2015MNRAS.450.2739M} and observations
(Table~\ref{tab:lxlgammaratiotable}).

The inferred properties of the shock producing the observed thermal X-rays 
are inconsistent with the inferred properties of the shock accelerating the GeV
$\gamma$-ray emitting particles. Earlier, \cite{2016MNRAS.463..394V} suggested 
the presence of different shock systems responsible for the thermal X-ray and non-thermal radio emission in novae.
The correlated optical and GeV variability seen in two novae by \cite{2017NatAs...1..697L}
and \cite{2020NatAs.tmp...79A}, together with GeV-to-optical flux ratios $\approx 0.01$, 
imply that the majority of GeV-emitting shock energy eventually emerges as
radiation. The shock should produce most of its thermal output in X-rays that
get absorbed and eventually escape as optical photons 
\citep[see Sec.~3.1 in ][]{2014MNRAS.442..713M}. X-rays from the GeV-emitting shock may be
completely hidden from our {\em NuSTAR} observations if the emission is sufficiently soft 
(k$T\lesssim0.6$\,keV corresponding to $v_{\rm shock} = 640$\,km\,s$^{-1}$) 
and if it disappears before {\em Swift} observations can probe 
these low energies (thanks to decreasing intrinsic absorption), after $t_0+63$\,d.

\subsection{Location of the X-ray emitting shock}
\label{sec:shocklocation}

The location of the X-ray emitting shock with respect to the binary system, 
the expanding nova ejecta and the $\gamma$-ray emitting shock is unclear. 
Here we discuss a few possibilities.

\cite{2014MNRAS.442..713M} consider the collision of nova ejecta with a dense external shell
and suggest the forward shock as the source of X-rays. 
There is no widely accepted observational evidence for the existence of dense circumbinary material in classical nova systems 
(with a dwarf donor; \citealt{2013AJ....145...19H, 2014ApJ...786...68H}, 
but see \citealt{2020MNRAS.494..743M}), however such material clearly exists 
in symbiotic systems (with giant donors; e.g., \citealt{1990ApJ...349..313S}). 

Optical spectroscopy has long indicated that, in many cases, nova outbursts produce multiple ejections 
with different velocities \citep[e.g.,][]{1944PA.....52..109M, 1966MNRAS.132..317F, 2011A&A...536A..97F, 2019arXiv190309232A}. 
Aydi et al. (2020b in prep.) argue that the presence of at least two physically distinct outflows is a common feature of novae.
Collision between a high-velocity wind that catches up with a low-velocity ejection launched earlier 
may produce the shock. 
A specific variation on the multiple outflows scenario was proposed by \cite{2014Natur.514..339C} based on radio imaging of 
the $\gamma$-ray-detected nova V959\,Mon. In this scenario, a slow equatorial outflow is launched at the time
of the nova outburst, and is later followed by a fast wind driven by the intense radiation from the white dwarf. 
The slow outflow is drawn from the puffed-up nova envelope that may remain gravitationally bound to the system, 
and gradually expands due to energy input from the binary system's orbital motion
\citep{2016MNRAS.461.2527P}, in analogy to the common envelope phase in binary system evolution \citep{1990ApJ...356..250L, 2013A&ARv..21...59I}. 
The shocks in this scenario form at the interface between the white dwarf wind and the equatorial outflow. 

One could also imagine that the accretion disc around a white dwarf could survive the nova explosion. 
The presence of accretion discs was reported 
in novae during the SSS phase \citep{2010AIPC.1248..197S,2011AAS...21733811W,2014ASPC..490..199M,2018MNRAS.480..572A}. 
The shock may form at the interface between the white dwarf wind and the accretion disk.
The orbital motion of the donor star within the expanded atmosphere of
the nuclear-burning white dwarf may create a bow shock. The expected orbital 
velocity $\sim200$\,km\,s$^{-1}$ is smaller than the shock velocity derived
from the plasma temperature (Eqn.~\ref{eq:ktshock}) and it should not
decrease with time (\S\,\ref{sec:thermalx}). 

Finally, it has been well established that radiation-driven winds of massive stars 
are clumpy \citep{2017SSRv..212...59M,2018A&A...611A..17S}. Interaction of
the dense clumps with the surrounding low-density wind and with each other
produce strong shocks \citep{1988ApJ...335..914O,1997A&A...322..878F}.
A hot nuclear-burning white dwarf should drive an intense wind. 
The combination of bound-free and line opacities is driving the wind of hot massive
stars, while in the even hotter novae the dominating opacity mechanism should be 
Thomson scattering \citep{2001MNRAS.326..126S}. At super-Eddington
luminosities found in novae, the Thomson scattering-supported wind should be
inhomogeneous, as it originates in an unstable atmosphere \citep{2001ApJ...549.1093S}.
Indeed, the ejecta of novae are well-known to be clumpy, as demonstrated by resolved imaging \citep{O'Brien_Bode_2008} 
and spectroscopic observations \citep[e.g.,][]{1994ApJ...426..279W,2013A&A...553A.123S, 2018ApJ...853...27M}.
Multiple shocks associated with multiple clumps distributed across the ejecta 
may give rise to nova X-ray emission \citep{2016JPhCS.728d2001W}---but in this case 
we might expect shocks at a range of absorbing columns, 
including some that are relatively unabsorbed even at early times.
\cite{2008ApJ...673.1067N} consider this scenario for the X-ray emission of RS~Oph with the
shocked clumps originating either in the nova ejecta or in the red giant
companion wind (red giant winds are reported to be clumpy;
\citealt{2006PhDT.........4C,2008ASPC..401..166E}; 
with maser observations indicating volume filling factors $<0.01$; \citealt{2012A&A...546A..16R}).

As we have an estimate of the shock velocity (\S\,\ref{sec:thermalx}), 
the X-ray variability time-scale, $t_{\rm var}$, can give us a clue 
about the size of the X-ray emitting region, $l = v_{\rm shock} t_{\rm var}$. 
The absence of variability on a $\sim100$\,ks time-scale (\S\,\ref{sec:nustarlc}) 
suggests, that on day 36 the X-ray emitting region was $\gtrsim$1~AU in size. 
Alternatively, the emitting region(s) might be small, but produce a stable flux of X-rays, 
or variations from a large number of emitting regions might average out \citep{1997A&A...322..878F}.

\subsection{The missing SSS phase}
\label{sec:missingsss}
A notable feature of \nova{} is the absence of a pronounced boundary between 
hard and super-soft X-ray emission observed in many other novae
\citep{2011ApJS..197...31S,2012BASI...40..353N,2013ApJ...768L..26P,2019arXiv190802004P}.
This is apparent from both the spectral fits 
(Table~\ref{tab:swiftmodels}, Fig.~\ref{fig:xrtspecspecev}) 
and from the absence of an abrupt increase in soft X-rays in Fig.~\ref{fig:swiftlc}.
Instead, the soft X-rays gradually brighten as the hard X-rays gradually fade.

The duration, temperature and luminosity of the SSS emission depend on
the mass of the white dwarf. Higher mass white dwarfs tend to produce brighter, hotter and 
more short-lived SSS compared to their low-mass counterparts
\citep{2011A&A...533A..52H,2013ApJ...777..136W,2017PhDT.......125W}.
The absence of SSS emission in \nova{} suggests that it either ended before
the nova ejecta became transparent to soft X-rays 
(a short-lived SSS implies a high mass of the white dwarf)
or that the SSS emission was 
sufficiently faint and soft to be completely hidden by the Galactic absorption (\S\,\ref{sec:extinction})
implying a low-mass white dwarf. 
The support the latter possibility comes from the CO composition of the white dwarf 
(\S~\ref{sec:ejectaabund}) together with the slow decline of 
the nova \citep{2005ApJ...623..398Y,2017ApJ...839..109S} and 
the slow ejecta velocities observed from the optical spectral lines 
\citep{1985ApJ...291..812K,1992A&A...262..487F,1994ApJ...437..802K} 
which are associated with a low-mass white dwarf. 

Another speculative possibility is that the shock-heated region may be so close to the white dwarf 
(i.e., the surviving accretion disc scenario in \S\,\ref{sec:shocklocation}) 
that shock energy may contribute to
heating the outer layers of the white dwarf. In this scenario, shocks and nuclear burning heat essentially
the same region of plasma near the surface of the white dwarf, blurring the boundary between the shock-powered and SSS emission.
The blackbody fits to the {\em Swift}/XRT spectra after
$t_0+250$\,d result in k$T<0.5$\,keV (Table~\ref{tab:swiftmodels}), 
i.e. qualifying as SSS emission (according to an observational definition; \S\,\ref{sec:xraynovaintro},~\ref{sec:thermalx}), 
but are still considerably hotter than the emission expected even from a very massive white dwarf \citep{2013ApJ...777..136W}. 

\section{Conclusions}
\label{sec:conclusions}

We conducted a joint analysis of {\em NuSTAR}, {\em XMM-Newton}, {\em Swift}
and {\em Fermi}/LAT observations of nova \nova{}. The observation 36 days after the explosion was only the second simultaneous {\em NuSTAR}/{\em Fermi} detection (out of four classical novae observed -- Table~\ref{tab:lxlgammaratiotable}). Our conclusions can be summarized as follows:

\begin{itemize}

 \item The X-ray emission of \nova{} in the {\em NuSTAR} band is soft (the photon index would be $\Gamma \simeq 4$; \S\,\ref{sec:nustarspec}),
and we attribute it to optically-thin thermal plasma of temperature k$T=$4--9\,keV 
(Table~\ref{tab:nustarmodels}).
We found no evidence for a non-thermal contribution to the 3.5--78\,keV emission (\S\,\ref{sec:nonthermalx}).

 \item The nova ejecta have highly non-solar abundances, consistent with ejection from the surface of a CO white dwarf (\S\,\ref{sec:ejectaabund}).

 \item \nova{} does not show distinct SSS emission, which may be an indication of a low white dwarf mass (\S\,\ref{sec:missingsss}). 
Instead, the X-ray spectral evolution of \nova{} can be described 
as a single optically-thin thermal emission component of gradually decreasing 
temperature hidden behind a column density that is also decreasing with time (\S\,\ref{sec:xrtspectroscopy}).

 \item The evolution of the absorbing column $N_\mathrm{H}$ with time
(Fig.~\ref{fig:nhtime}) implies that $5\times10^{-5}$\,M$_\odot$ 
(corrected for heavy element abundances; \S\,\ref{sec:ejectamass}) were ejected 24 days after the start of the eruption.
Gradual acceleration of the ejecta and ejection over a prolonged period of
time are the alternative to the late ejection scenario.

 \item The absence of variability on $\lesssim100$\,ks time-scale in the {\em NuSTAR} band suggests
that the X-ray emitting region is larger than $\sim$1~AU (\S\,\ref{sec:shocklocation}).

 \item Contrary to theoretical expectations,
the {\it thermal} hard X-ray emission observed by {\em NuSTAR} is much fainter (in $\nu F_{\nu}$ units)
than the simultaneous GeV $\gamma$-ray emission ($L_{\rm 20\,keV}/L_{\rm 100\,MeV} = 0.02$; \S\,\ref{sec:lxlg}). 
\nova{} is the fourth $\gamma$-ray emitting nova to demonstrate such low $L_{X}/L_{\gamma}$ (Table~\ref{tab:lxlgammaratiotable}). 
The low X-ray luminosity may indicate that the shocks responsible for the X-ray emission are not the same 
as the ones accelerating GeV-emitting particles, that X-rays are suppressed in nova radiative shocks, 
that particle acceleration is surprising efficient, and/or  that the radiative shock approximation 
is not applicable to these shocks (\S\,\ref{sec:thermalx}). 

 \item The {\it non-thermal} hard X-ray emission contribution is
constrained at $L_{\rm 20\,keV}/L_{\rm 100\,MeV} < 5 \times 10^{-4}$.
This rules out leptonic models of the nova GeV emission (\S\,\ref{sec:nonthermalx}).

\end{itemize}

Future Very Long Baseline Interferometry (VLBI) radio observations may image synchrotron emission of the accelerated particles and 
pinpoint the location and geometry of $\gamma$-ray and X-ray emitting shocks (however, free-free absorption and synchrotron
self-absorption may hamper early radio imaging). 
Simultaneous GeV and hard X-ray observations of future novae may provide further insights into the ``missing X-rays'' problem (both thermal and non-thermal).
The ultimate proof for hadronic mechanism of the GeV emission will be detection of neutrino emission from a nova.

\section*{Data availability}

The processed data underlying this work are available at the request to 
the first author and here\footnote{\url{http://scan.sai.msu.ru/~kirx/data/V906_Car__public_data/}}. 
The raw data are publicly available at {\em NuSTAR}, 
{\em XMM-Newton}, {\em Swift} and {\em Fermi} science archives.

\section*{Acknowledgements}
KVS thanks 
Dr.~Victoria Grinberg for drawing attention to the problem of clumpy stellar winds
and
Dr.~Marina Orio for pointing out the report of possible X-ray fireball detection. 
We thank the anonymous referee for the very useful comments. 
This material is based upon work supported by the National Science Foundation under Grant~No.~AST-1751874.
We acknowledge support for this work from NASA grants Fermi/80NSSC18K1746 and NuSTAR/80NSSC19K0522 and from 
a Cottrell Scholarship from the Research Corporation.
KLP acknowledges support from the UK Space Agency. JS acknowledges support from the Packard Foundation. 
KLL is supported by the Ministry of Science and Technology of the Republic of China (Taiwan) through grant 108-2112-M-007-025-MY3. 
IV acknowledges support by the Estonian Research Council grants IUT26-2 and IUT40-2, and by the European Regional Development Fund (TK133). 
%




\bibliographystyle{mnras}
\bibliography{asassn18fv} 





\appendix

\section{Estimates of inner/outer shell velocity ratio for the ``Hubble flow'' model of nova ejecta}
\label{sec:vminvmaxliterature}

Estimates of $v_{\rm min}/v_{\rm max}$ ratio can be derived from modelling the multi-frequency 
radio lightcurve of thermal emission from a nova (see \citealt{1979AJ.....84.1619H, Seaquist_Bode_2008}; 
\citealt{2016PhDT........30W} and appendix~A in \citealt{2018ApJ...852..108F}). 
\cite{2014ApJ...792...57R} discussed the effects of the expected bipolar (dumbbell-shaped) geometry 
of the ejecta on the results of fitting the radio lightcurves with the simple spherically symmetric ``Hubble flow'' model. 
They found that the spherical model fits overpredict the ejecta mass by up to a factor of two and underpredict 
the shell thickness (proportional to the $v_{\rm min}/v_{\rm max}$ ratio) by up to an order of magnitude 
(so $v_{\rm min}$\,should be close to $v_{\rm max}$). 
The magnitude of the discrepancy between the spherical and bipolar models depends mostly on 
the departure from spherical symmetry and only weakly on inclination. 
The clumpiness of the nova ejecta is an additional source of uncertainty in modelling \citep{2012BASI...40..293R}.
Shocks within the nova ejecta producing thermal \citep{2014MNRAS.442..713M} 
and synchrotron \citep{2016MNRAS.463..394V,2019arXiv190801700S} emission may further complicate modelling 
of nova radio lightcurves. Table~\ref{tab:vminvmax} summarizes the $v_{\rm min}/v_{\rm max}$
values reported in the literature.

\begin{table}
        \centering
        \caption{The inner/outer shell velocity ratios from the literature}
        \label{tab:vminvmax}
        \begin{tabular}{c l c} 
                \hline
Nova      & $v_{\rm min}/v_{\rm max}$  & Reference  \\
                \hline
V1324\,Sco & $0.447_{-0.079}^{+0.10}$   & \cite{2018ApJ...852..108F} \\ 
V959\,Mon  & $0.083$                    & \cite{2014Natur.514..339C} \\ 
V5589\,Sgr & $0.84$                     & \cite{2016MNRAS.460.2687W} \\ 
V1723\,Aql & $0.17$                     & \cite{2016MNRAS.457..887W} \\ 
T\,Pyx     & $0.25$                     & \cite{2014ApJ...785...78N} \\ 
V723\,Cas  & $0.24 \pm 0.1$             & \cite{2005MNRAS.362..469H} \\
V1974\,Cyg & $0.46^a$                   & \cite{1996ASPC...93..174H} \\
V351\,Pup  & $0.74^b$                   & \cite{2017ApJ...840..110W} \\
V838\,Her  & $0.042$                    & \cite{1996ASPC...93..174H} \\
V827\,Her  & $0.25$                     & \cite{1996ASPC...93..174H} \\
V1819\,Cyg & $0.2$                      & \cite{1996ASPC...93..174H} \\
QU\,Vul    & $0.87$                     & \cite{1996ASPC...93..174H} \\
V1500\,Cyg & $0.036$                    & \cite{1979AJ.....84.1619H} \\
FH\,Ser    & $0.048$                    & \cite{1979AJ.....84.1619H} \\
HR\,Del    & $0.44$                     & \cite{1979AJ.....84.1619H} \\
                \hline
        \end{tabular}
\begin{flushleft}
$^a$~\cite{1993MNRAS.263L..43I} report $v_{\rm min}/v_{\rm max} = 0.16$.\\
$^b$~\cite{1996ASPC...93..174H} found $v_{\rm min}/v_{\rm max} = 0.069$.\\
{\bf Column designation:}
Col.~1~-- Nova name;
Col.~2~-- The velocity ratio derived from the ``Hubble flow'' model;
Col.~3~-- Citation.
\end{flushleft}
\end{table}


\bsp 
\label{lastpage}
\end{document}